\documentclass[prx,aps,floatfix,10pt,twocolumn,nofootinbib]{revtex4-2}

\usepackage[T1]{fontenc}
\usepackage[utf8]{inputenc}
\usepackage{microtype}
\usepackage{array}
\usepackage{graphicx}
\usepackage{dcolumn}
\usepackage{bm}
\usepackage{amsmath}
\usepackage{amsfonts}
\usepackage{amssymb}
\usepackage{amstext}   
\usepackage{amsthm}
\usepackage{array}
\usepackage[english]{babel}
\usepackage{makecell}
\usepackage{dsfont}
\usepackage{nameref}
\usepackage{color}
\usepackage{float}
\usepackage[outdir=./]{epstopdf}
\usepackage{nameref}
\usepackage{subfloat}
\usepackage[colorlinks = true, citecolor = blue, urlcolor =cyan]{hyperref}
\usepackage[figure]{hypcap}
\usepackage{multirow}
\usepackage{makecell}
\usepackage{xcolor}
\usepackage{mathtools}
\usepackage{mathdots}
\usepackage[notransparent]{svg}
\usepackage{todonotes} %I like to use this one to highlight missing bits for myself. -Alicia
\usepackage[symbol]{footmisc}
\newcolumntype{C}{>{$}c<{$}}
\AtBeginDocument{
\heavyrulewidth=.08em
\lightrulewidth=.05em
\cmidrulewidth=.03em
\belowrulesep=.65ex
\belowbottomsep=0pt
\aboverulesep=.4ex
\abovetopsep=0pt
\cmidrulesep=\doublerulesep
\cmidrulekern=.5em
\defaultaddspace=.5em
\tabcolsep=7pt
}
\usepackage{booktabs}
\usepackage{verbatim}

\definecolor{emerald}{rgb}{0.07, 0.53, 0.03}

\usepackage{braket}

\usepackage{amsbsy}

\usepackage{nicefrac} %added for equations

\usepackage{soul}   % SOUL Added for underlines
\setstcolor{red} % strikethrough color
\setul{}{1.5pt}  % line thickness

\newcommand{\ak}[1]{\textbf{\color{blue}[AK: #1]}}

\newcommand{\five}{Q2}
\newcommand{\eight}{Q3}
\newcommand{\sixteen}{Q1}
\begin{document}

%\title{A Superconducting-Circuit Waveguide QED System with Unconventional Band Structures}
%\title{A Multimode Circuit-QED System with Unconventional Band Structures for Photon-Mediated Spin Models}
\title{A Circuit-QED Lattice System with Flexible Connectivity and Gapped Flat Bands for Photon-Mediated Spin Models}

\author{Kellen O'Brien}
\thanks{Contact Author: kobrie@umd.edu}
\affiliation{Department of Physics and JQI, University of Maryland, College Park, MD 20742, USA}

\author{Maya Amouzegar}
\affiliation{Department of Physics and JQI, University of Maryland, College Park, MD 20742, USA}

\author{Won Chan Lee}
\affiliation{Department of Physics and JQI, University of Maryland, College Park, MD 20742, USA}

\author{Martin Ritter}
\affiliation{Department of Physics and JQI, University of Maryland, College Park, MD 20742, USA}

\author{Alicia J.  Koll\'ar}
\affiliation{Department of Physics, JQI, and QMC, University of Maryland, College Park, MD 20742, USA}

\preprint{APS/123-QED}

\date{December 2, 2025}% It is always \today, today,
             %  but any date may be explicitly specified

\begin{abstract}
%Quantum spin models underlie a wide variety of solid-state phenomena, but outside of a few special limits, classical simulation of them remains extremely challenging.
%Quantum spin models underlie a wide variety of solid-state phenomena, but outside of select special limits, classical simulation of these models is extremely challenging. 
Quantum spin models are ubiquitous in solid-state physics, but classical simulation of them remains extremely challenging.
Experimental testbed systems with a variety of spin-spin interactions and measurement channels are therefore needed.
One promising potential route to such testbeds is provided by microwave-photon-mediated interactions between superconducting qubits, where native strong light-matter coupling enables significant interactions even for virtual-photon-mediated processes. In this approach, the spin-model connectivity is set by the photonic mode structure, rather than the spatial structure of the qubit.
%and where the spin-model connectivity is set by the photonic mode structure, rather than the shape of the qubit. 
Lattices of coplanar-waveguide (CPW) resonators have been demonstrated to allow extremely flexible connectivities and can therefore host a huge variety of photon-mediated spin models. However, large-scale CPW lattices with non-trivial band structures have never before been successfully combined with superconducting qubits.
Here we present the first such device featuring a quasi-1D CPW lattice with multiple transmon qubits.
We demonstrate that superconducting-qubit readout and diagnostic techniques can be generalized to this highly multimode environment and observe the effective qubit-qubit interaction mediated by the bands of the resonator lattice.
This device completes the toolkit needed to realize CPW lattices with qubits in one or two Euclidean dimensions, or negatively-curved hyperbolic space, and paves the way to driven-dissipative spin models with a large variety of connectivities.
%\wcl{mode-mode spec story would be good to be included... Maybe can split into two sentences: 1. readout 2. Observation of effective qubit-qubit interaction. Should I modify?}\ak{I think it's too specific for the abstract.}
%a quasi-1D CPW resonator lattice with three coupled transmon qubits.

\end{abstract}

%\keywords{Suggested keywords}%Use showkeys class option if keyword
                              %display desired
\maketitle

%\tableofcontents

%%%%%%%%%%%%%%%%%
\section{Introduction}\label{sec:intro}

%%%%%%%%%%%%%%%%%%%%%%%%%%%%%%%%%%%
%Implementing and simulating quantum spin models is one of the forefronts of current physics research, both theoretical and experimental. 

Quantum spin models underlie a wide variety of solid-state phenomena from ferromagnetism, to spin glasses, to parts of the phase diagram of high-temperature superconductors \cite{girvin,SpinGlass,AndersonSpinLiquid}. Efficient classical simulation algorithms exist in special types of connectivity, but the general phenomenology of these systems remains an outstanding grand challenge, particularly when a spin Hamiltonian is accompanied by drive and dissipation. Implementing these models in synthetic quantum systems, which offer access to novel local or dynamical measurements, holds the potential to provide new insights into this challenge. 
%
% Here we present an experimental platform, using superconducting qubits coupled to microwave resonator arrays, capable of implementing spin-spin interactions with a large variety of connectivities, spanning from simple 1D chains, to 2D, to positively-curved spherical spin models, and even to hyperbolic spin models \cite{Houck_Nature_QS,bienias:Hyperbolic}.
% \ak{underdevelopment}
% previous
% - demonstrated flexible resonator connectivity
% - we show that can combine that kind of array with qubits, and not mess up
% - meaning that the previous resonator connectivity becaomes qubit connectivity
% - we complete the toolkit.
Here we present an experimental platform suitable for implementing spin-spin interactions with a large variety of connectivities, spanning from simple 1D chains, to 2D, to positively-curved spherical spin models \cite{Houck_Nature_QS,bienias:Hyperbolic}, and even to hyperbolic spin models, by leveraging 
the flexible connectivity of microwave resonator arrays~\cite{Kollar:2019hyperbolic,Kollar:2019linegraph}.
We demonstrate integration of transmon qubits without compromising the versatility of the resonator array, as well as qubit-qubit interactions mediated by the bands of the array, 
completing the experimental toolkit needed to realize two-dimensional or hyperbolic spin models.

The crucial ingredient underlying the flexibility of this platform is the use of microwave photons to mediate interactions between effective spin-$1/2$ degrees of freedom formed by qubits. Direct qubit-qubit coupling inherently gives rise to a spin-spin interaction which follows the geometry of the device. By contrast, using a photon to mediate coupling allows the geometry of the spin-spin interaction to be decoupled from the physical locations of the qubits, and in these cases, the limits on achievable interactions are set by the ability to control and modify the propagation of the mediating photon. In the simplest case, a single-mode cavity mediates an all-to-all interaction between qubits, leading to trivial connectivity.
An interaction that generates a spin model with non-trivial connectivity requires the presence of multiple modes in a bandwidth set by the qubit-photon interaction strength \cite{Douglas_2015, bienias:Hyperbolic, Sundaresan, Benedetto2024}. 
In such cases, the dispersion relation and spatial profiles of the photon modes determine the interaction profile.

%an effective spin-spin interaction with non-trivial spatial structure. 
%multiple modes in the bandwidth set by the qubit-photon interaction strength.
%A spatially non-trivial interaction requires the presence of multiple photon modes.

The circuit QED platform \cite{Blais:revmodphys}, featuring superconducting qubits coupled to microwave resonators, is an ideal host for such models due to the combination of two types of hardware properties. 
(i) Superconducting microwave resonators have very high quality factors and can be implemented in a wide variety of different ways, with highly variable form factors on chip. As a result of this flexibility, superconducting microwave resonator arrays have been shown to host linear, quadratic, and flat bands in both one and two dimensions~\cite{Underwood:imaging, Houck:earlylattice, Houck_Nature_QS, Kollar:2019linegraph, Painter_Markov, Painter_SSH, Painter_2023, Martinez_2023, Simone:resonatorarray}, and non-Euclidean structures \cite{Kollar:2019hyperbolic}. Flat bands are particularly important since they quench kinetic energy, permitting interaction-driven strongly correlated phases to emerge in many-body systems. Furthermore, beyond their intrinsic interest, many-body systems on manifolds with spatial curvature enable experimental access to exotic topological observables. (ii) Superconducting qubits are frequency tunable, and natively exhibit strong light-matter coupling, which can easily be adjusted by orders of magnitude. The combination of these two properties gives unprecedented ability to tailor the spatial structure of strong qubit-qubit interactions, without the loss in the photon modes significantly damaging native qubit coherence times.
%For example, Ref.~\cite{Kollar:2019hyperbolic} showed that coupled coplanar-waveguide resonators can implement a tight-binding model which is effectively hyperbolic, which in turn will mediate a spin model with hyperbolic connectivity if qubits are incorporated into the lattice \cite{bienias:Hyperbolic}.
%Refs.~\cite{Kollar:2019hyperbolic,Kollar:2019linegraph} shows that resonators of this kind also provide 
%, and thus the effective connectivity of an effective spin-model, while minimizing the 

Multimode cavity-QED models with the potential to host boson-mediated spin models with non-trivial connectivity have also been studied in atomic systems with multiple discrete modes \cite{kollar2017supermode, Papagreorge2016, Kollar:apparatus, Vaidya:2018fp, Gopalakrishnan09, Gopalakrishnan10, Gopalakrishnan:2011jx, Gopalakrishnan:2012cf, Guo:2019ej, Guo:2019fr, Marsh2024, Vuletic:2001ur, vuletic:prl, schine:landau, schine:FloquetPolariton, schine:Laughlin}; or a continuum of modes from a waveguide  \cite{RevModPhys_Waveguides, Fayard_2021}, a photonic crystal \cite{Douglas_2015}, or a BEC \cite{Krinner:2018, Stewart:2020, Lanuza:2022}; or with Raman coupling in a single-mode cavity \cite{Periwal_2021}.
%\cite{SchleierSmith:2010fs, SchleierSmith:2010jg, Leroux:2010kl}.
%in which periodic modulation of a medium is used to tailor photon dispersion relations \cite{Douglas_2015}. 
However, none of these systems can achieve the large coupling strengths and the variety of connectivities possible in superconducting microwave circuits.

% \ak{Tweaked Won's previous version of this paragraph 3-6-25. Let me know what you guys think.} \wcl{It looks better!}
Previous multimode QED and waveguide QED experiments with superconducting circuits have used modulated waveguides \cite{Liu, Sundaresan}, simple one-dimensional chains of resonators \cite{vrajitoarea2024, Simone:resonatorarray, Painter_SSH, Painter_Markov, Painter_2023}, or 2D Hofstadter lattices \cite{Owens_2018, Schuster:Chiralcavity}. In these cases, the attainable dispersion relations and spatial profiles of the photon modes have been limited by either the simplicity of the waveguide chain or by the rigid form factor of the 3D cavities.
In contrast, coplanar-waveguide (CPW) resonators \cite{Goppl_2008} can form resonator arrays \cite{Houck_Nature_QS, Houck:earlylattice, Koch:AnnPhysBerl, Underwood:imaging} with low-disorder \cite{Houck:earlylattice} and highly flexible connectivities \cite{Kollar:2019hyperbolic, Kollar:2019linegraph}. 
%It has been shown that this technique can produce lattices in which the photons propagate in an effective negatively-curved hyperbolic space \cite{Kollar:2019hyperbolic}, and that these lattice systems have a strong tendency to produce flat bands \cite{Kollar:2019hyperbolic, Kollar:2019linegraph}. It has also been shown theoretically that the photon-mediated qubit-qubit interaction that a hyperbolic CPW lattice would produce also follows the hyperbolic metric \cite{bienias:Hyperbolic}. Furthermore, the flat bands exhibited by these lattices consist entirely of wave functions which flip sign on neighboring sites \cite{Kollar:2019hyperbolic, Kollar:2019linegraph}, and which can therefore give rise to short-range frustrated magnetic interaction between qubits.
It has been demonstrated that this technique can produce lattices in which the photons propagate in an effective negatively-curved hyperbolic space \cite{Kollar:2019hyperbolic}, and it has been shown theoretically  that the photon-mediated qubit-qubit interaction that a hyperbolic CPW lattice would produce also follows the hyperbolic metric \cite{bienias:Hyperbolic}. In general, CPW resonator lattices also have a strong tendency to produce flat bands that consist entirely of wave functions which flip sign on neighboring sites \cite{Kollar:2019hyperbolic, Kollar:2019linegraph}, and which can therefore give rise to short-range frustrated magnetic interaction between qubits.
However, previous experimental studies of coplanar-waveguide arrays have either been limited to simple 1D chains with many qubits but large disorder \cite{Fitzpatrick_2017}, or higher-dimensional connectivity and low disorder but no qubits \cite{Underwood:imaging, Houck:earlylattice, Kollar:2019hyperbolic}.

Here we present the first CPW lattice device featuring both a large-scale low-disorder resonator array with non-trivial band structure and multiple superconducting qubits. 
Our device features a Euclidean quasi-one-dimensional microwave resonator lattice with a non-trivial unit cell, which gives rise to conventional quadratic bands, linearly dispersing bands, and flat bands with localized eigenstates.
We incorporate three flux-tunable transmon qubits directly into the CPW lattice and show that the low-disorder microwave resonator array survives the addition of the qubits and their associated flux-bias control.
%We show that the qubits can be incorporated into the resonator network without significantly impacting the flexible connectivity of the resonator array. 

We demonstrate that conventional superconducting-qubit readout and diagnostic techniques \cite{Blais:revmodphys}, though originally developed for single-mode systems, can be applied in this highly multimode environment, and can be used to observe the effective qubit-qubit interaction mediated by the bands of the resonator lattice. 
Additionally, we extend prior non-linear spectroscopy techniques in superconducting circuits, such as Kerr spectroscopy \cite{Bosman_2017,Peugeot_2024} and qubit two-tone spectroscopy \cite{Blais:revmodphys}, to implement reliable in-situ measurements of the mode spectrum of the CPW lattice. The technique, which we term mode-mode spectroscopy, leverages the presence of qubits to suppress the effects of background modes and does not require two-port transmission of all modes, enabling observation of localized modes as well as delocalized ones.
Previous related techniques were either extremely time-consuming \cite{Houck:earlylattice} or highly sensitive to the location of probe ports and unable to distinguish lattice modes from some types of parasitic modes induced by device packaging \cite{Kollar:2019hyperbolic}. 
%we develop a novel in situ diagnostic technique, which leverages the presence of qubits, to measure the mode spectrum of the CPW lattice.
%Previous related techniques were either extremely time-consuming \cite{Houck:earlylattice} or highly sensitive to the location of probe ports and unable to distinguish lattice modes from some types of parasitic modes induced by device packaging \cite{Kollar:2019hyperbolic}. 
%In contrast, our method is fast, shows very strong suppression of the effects of background modes, and does not require two-port transmission of all modes, enabling observation of localized modes, as well as delocalized ones.
The device and measurement techniques presented here pave the way to CPW-lattice-based spin models with frustrated, 2D, and even hyperbolic connectivity.

The remainder of the manuscript is organized as follows. Section \ref{sec:background} provides the theoretical description for a general multimode cavity-QED system. Section \ref{sec:circuitQED} then introduces the more specific case of transmon qubits coupled to a CPW lattice. Section \ref{sec:device} introduces the band structure of our chosen quasi-one-dimensional lattice and shows how the device is implemented using superconducting-circuit hardware. In Section \ref{sec:bandcharacterization} we present characterization of the mode spectrum of the device, both through standard transmission spectroscopy and via our newly developed measurement technique. Lastly, Section \ref{sec:qubitmeasurements} demonstrates the application of standard qubit readout in a CPW lattice and shows a tunable qubit-qubit interaction mediated by the photonic modes of the resonator lattice.

%%%%%%%%%%%%%%%%%%%%%
\section{Background}\label{sec:background}
Traditional cavity QED describes an interaction between a single qubit/atom/spin and a single photon mode.
Extensions to a continuum of photon modes have been studied in the context of emitters coupled to a waveguide, known as waveguide QED \cite{RevModPhys_Waveguides, Fayard_2021}.
CPW lattices produce a large number of discrete modes which approximate the continuous density of states of the bands of an infinite lattice. These systems therefore approximate the full continuum of waveguide QED.
To simplify the language and highlight the commonalities between different ways of creating multimode systems, we refer to all systems in which an effective continuum of highly-coupled modes with a dispersion relation is relevant to forming photon-mediated interactions as (generalized) photonic crystal and waveguide QED systems.
In the remainder of this section, we present a general framework for describing multimode QED devices of this kind with multiple qubits.
Characteristics of these models which are specific to superconducting circuits and CPW resonators, in particular, will be discussed in Section \ref{sec:circuitQED}.
%Waveguide QED, on the other hand, studies quantum emitters coupled through a continuum of many photonic modes \cite{RevModPhys_Waveguides, Fayard_2021}. The atom-photon interaction strength becomes stronger compared to free space in both cases. However, the inherent geometry of the waveguide QED system enables an efficient detection of the photon and interaction by mediating photon modes. In this section, we will describe the general framework for a multi-qubit waveguide QED device of this kind, before presenting a description of the particular hardware implementation in Sec.~\ref{sec:circuitQED}.

\subsection{Cavity and Waveguide QED}\label{subsec:waveguide QED}
The simplest Hamiltonian for a cavity-qubit coupled system is known as the Jaynes-Cummings Hamiltonian \cite{Kimble:2005wd,Carmichael:LesHouches,wallsMilburn}:
\begin{equation}\label{eqn:JC}
    H_{JC} = \frac{1}{2} \hbar \omega_q \sigma_z + \hbar \omega_c (a^\dagger a + 1/2) +  \hbar (g^* \sigma^+ a + g \sigma^- a^\dagger),
\end{equation}
where $\omega_q$ is the qubit (atomic) transition frequency, $\omega_c$ is the cavity frequency, and $g$ is the light-matter interaction strength. The properties of such a system are controlled by the interplay between the coherent coupling $g$ and the native decay rates of the system: $\gamma$, the qubit decay rate, and $\kappa$, the cavity decay rate.

% \ak{transitions need reworking}
% In this section, we will describe the general framework for a multi-qubit waveguide QED device. The most conventional case is a qubit coupled to a waveguide with a fully continuous spectrum \cite{XYZ}. 
In this work we consider systems where the waveguide is formed by a resonator network. 
We therefore present the general discrete-mode case of waveguide QED here, but extension to a full continuum is straightforward \cite{cohentanoudji}.
In this framework the qubit and cavity portions of the Hamiltonian become
\begin{equation}
\label{eqn:multiqubit}
    H_{qubit} =  \sum_{j} {\frac{\hbar \omega_j}{2} \sigma_{j,z} }
\end{equation}
and 
\begin{equation}
\label{eqn:multicav}
    H_{photon} = \sum_{k} \hbar \omega_k a_k^{\dag} a_k,
\end{equation}
where $\sigma_{j,z}$ and $\omega_j$ denote the Pauli $z$ operator and frequency of the $j^{th}$ qubit respectively, and we label the independent photon modes by a quantum number $k$. 
%The label $k$ indicates the momentum of the photon mode. A conventional one-dimensional waveguide studied in archetypical waveguide QED systems has a photon spectrum which is fully continuous, and Eq.~\ref{eqn:multicav} is more properly written as an integral.
The label $k$ indicates the momentum of the photon mode, but in general, this quantum number may also contain additional information such as a band index, and may refer to a standing wave mode with momentum $k$, rather than an ideal plane-wave solution. %, such as would be found with periodic boundary conditions.  

The Jaynes-Cummings interaction term generalizes to
\begin{equation}\label{eqn:multimodeJCint}
H_{int} =     \sum_{k,j}  \hbar\left(g^*_{j,k} \  a_k^{\dag} \sigma_{j,-} +  g_{j,k} \   a_k \sigma_{j,+} \right)\!,
\end{equation}
where the spatial structure of the photon modes and the location of the qubit is encoded in the values of the mode-dependent coupling constants $g_{j,k}$. For a delta-function-like atom $g_{j,k}$ has a simple form, namely the product of an overall coupling rate $g_0$ and the wave function of the photonic mode $\psi_k$ evaluated at the location of the qubit: $g_{j,k} = g_0 \, \psi_k(x_j)$. With this form of the coupling the interaction simplifies to 
\begin{equation}\label{eqn:multimodeJCint2}
H_{int} =      \hbar g_0 \sum_{k,j}  \left( \psi^*_k(x_j) a_k^{\dag} \sigma_{j,-} + \mbox{ h.c.} \right)\!.
\end{equation}
Any process which is second order in this Hamiltonian will give rise to an exchange interaction between qubits mediated by emission and reabsorption of photons. If the qubits couple resonantly to photon modes, then the qubits will directly inherit the ``flying'' character of photons and the qubit lifetime will generally decrease. On the other hand, if the photon modes are strongly coupled to the qubit but off-resonant, then direct hybridization is minimized and the photons mediating the exchange interaction will be virtual. This regime allows a significant photon-mediated qubit-qubit interaction without an immediate loss in qubit coherence. However, this involves creating a multimode photonic system with frequency regions in which there are no photon modes, known as photonic band gaps.

\subsection{Photon Bound States and Photon-Mediated Interactions}\label{subsec:boundstates}

Photonic band gaps \cite{John_Wang} are typically created through some form of periodic modulation of the medium. One type of device architecture involves periodic modulation of the index of refraction, for the case of optical photons \cite{Emanuel:revmodphys}, or the impedance \cite{Liu,Sundaresan}, for microwave photons, and generally has a full continuum of delocalized modes. Another class of devices instead uses periodic arrays of localized sites
%, either atoms \cite{Douglas_2015, Painter_Mirrors} or 
such as optical or microwave resonators \cite{Houck:earlylattice, Houck_Nature_QS, Kollar:2019hyperbolic, Simone:resonatorarray, Owens_2018, Schuster:Chiralcavity, Painter_Markov, Painter_SSH, Martinez_2023, Arrangoiz_Arriola_2018}. In these devices, the photonic modes are delocalized only by the coupling between the sites, and can be thought of as akin to tight-binding models of solids \cite{girvin}. 
The device presented in this work is chosen to be of the latter kind because the tight localization of the mode in the individual resonators leads to separation between the hardware properties that set the on-site energy and the properties that control the coupling, or hopping, between sites. 
%This separation of constraints has been shown to enable on-chip realization of a large variety of unconventional and flat-band lattices \cite{Kollar:2019hyperbolic, Kollar:2019linegraph}.
%\ak{Do we want to go here at this stage?} \ko{Feels a little out of place, think focus of paragraph should remain on bandgap creation methods}

% Regardless of how it is created, the existence of a band edge, where the photonic band gap ends and the photonic modes begin, induces a set of generic consequences, one of which is that the introduction of a qubit induces a so-called photonic bound state \cite{John_Wang, Douglas_2015}, where a localized mode with photonic character emerges, even though all of the photon modes are delocalized in the absence of coupling to the qubit. 
In the context of photonic band gaps, photon-mediated qubit-qubit interactions are often described as arising from the formation of photonic bound states \cite{John_Wang, Douglas_2015,Calaj_2016}.
These are localized modes with photonic character which emerge even if all of the photon modes are delocalized in the absence of coupling to the qubit. 
% Regardless of how it is created, the existence of a band edge, where the photonic band gap ends and the photonic modes begin, induces a set of generic consequences, one of which is that the introduction of a qubit induces a so-called photonic bound state \cite{John_Wang, Douglas_2015}, where a localized mode with photonic character emerges, even though all of the photon modes are delocalized in the absence of coupling to the qubit. 
If the qubit is in the band gap, there is an intuitive picture of the formation of the bound state in which the qubit hybridizes with the photon modes and acquires a cloud of virtual photons surrounding it. In the conventional case, where the dispersion of the photon modes above the band edge is quadratic ($\omega_k = \omega_{edge} + \beta k^2$) and the coupling constants $g_{j,k}$ are approximately equal, this photon cloud is known to take the form of an exponentially localized envelope, $e^{-\vert x \vert /L}$ \cite{John_Wang, Douglas_2015,Calaj_2016}, where $\vert x \vert$ is the distance from the qubit, and where the localization length $L$ is determined by both the band curvature $\beta$, and the detuning between the qubit and the band edge. 
The simple picture of the photonic bound state being a cloud of virtual photons surrounding a qubit is valid only for a qubit well inside the band gap. However, an exact solution of similar form exists at \emph{any} detuning from an ideal quadratic band edge and has been derived in Refs.~\cite{John_Wang,Liu, Sundaresan,Calaj_2016}.
If two qubits are coupled to the band edge, then interactions between them can be thought of as arising from the tail of the photonic bound state localized around one qubit driving transitions in the other qubit. Hence, the spatial structure of the photonic bound state dictates the spatial structure of the interaction.

%\ko{switch ALL of Chang underscore 2012}
% \ak{check if the general solution is only in Sundaresan, or if it's in the other ones too} \ko{Only in Sundaresan}
Alternatively, the photon-mediated qubit-qubit interactions can be modeled directly as a photon-exchange process \cite{Douglas_2015, Sundaresan, bienias:Hyperbolic}, as described in Section~\ref{subsec:waveguide QED}. 
For arbitrary detunings, the frequency of the photonic bound state and the resulting interaction between the $j^{th}$ qubit and the $l^{th}$ qubit must be found self-consistently \cite{Sundaresan,bienias:Hyperbolic,Calaj_2016}. However, if both qubits are in the band gap, then second-order perturbation theory can be used to obtain a simple form valid for any dispersion
\begin{equation}\label{eq:twoqint}
    H_{j,\ell} = \hbar \sigma_{j, -} \sigma_{\ell, +} \frac{1}{N}\sum_k \frac{g_0^2}{\Delta_k} \psi_k\left(x_j\right)\psi_k^*\left(x_\ell\right) + h.c. ,
\end{equation}
where $\sigma_{j, -}$ and $\sigma_{j, +}$ indicate lowering and raising operators on the $j^{th}$ qubit, and $\Delta_k$ is the detuning between the qubits and mode $k$ (assumed to be approximately equal for the two qubits). 
A quadratic band edge in a one-dimensional set of modes gives a simple  exponential interaction, but this can be changed by going to higher dimensions, non-quadratic dispersions \cite{Yidan:2022}, or non-plane-wave modes.
\section{Photonic Crystals in Circuit QED}\label{sec:circuitQED}

Since the form of the qubit-qubit interaction mediated by a photonic bound state is determined by the dispersion and mode functions of the band, creating spin models with a large variety of connectivities requires the ability to engineer diverse band structures with strong qubit-photon coupling. In this section we describe the hardware aspects of CPW lattices that make them ideally suited to this task, as well as the measurement techniques native to this microwave-frequency implementation of waveguide QED. We emphasize that the methods developed for single-mode circuit QED \cite{Blais:revmodphys} can be generalized to a full toolkit for design, characterization, and measurement of multimode waveguide-QED devices made with CPW resonators.

% The previous section discussed how photonic energy bands facilitate qubit-qubit interactions through the creation of photonic bound states. In this section we describe the hardware aspects of CPW lattices that make the creation of diverse band structures possible as well as the measurement techniques native to this microwave-frequency implementation of waveguide QED. We emphasize that the methods developed for single-mode circuit QED \cite{Blais:revmodphys} can be generalized to a full toolkit for design, characterization, and measurement of multimode waveguide-QED devices made with CPW resonators.

% Before presenting the details of our particular device geometry and measurement results in Sections~\ref{sec:device}-\ref{sec:qubitmeasurements}, we first summarize the underlying hardware building blocks and measurement techniques particular to circuit QED and CPW lattices, emphasizing how the methods that were originally developed for single-mode circuit QED \cite{Blais:revmodphys} can be generalized to a full toolkit for design, characterization, and measurement of multimode waveguide-QED devices made with CPW resonators.

% \subsection{Transmon Qubits and the Strong Dispersive Regime}\label{subsec:transmon}
% \subsection{Transmon Qubits and Two-Tone Spectroscopy}\label{subsec:spec_theory}
\subsection{Transmon Qubits and Readout}\label{subsec:transmon_and_readout}

The transmon is the most common superconducting qubit currently in use \cite{Koch:transmon,Blais:revmodphys}. It consists of an anharmonic oscillator
formed by a capacitor and a Josephson junction in parallel. The non-linearity of the Josephson junction \cite{Tinkham:book} introduces a negative anharmonicity, which results in the $\ket{0}\rightarrow \ket{1}$ transition frequency of the transmon, $\omega_q$, being higher in energy than all other transitions, allowing the transmon to be treated as an effective two-level system.
If the single Josephson junction is replaced with a DC SQUID \cite{Tinkham:book}, the transition frequency of the transmon becomes tunable in-situ via applied magnetic flux. 
Unlike atomic systems, which can typically only be tuned a few tens of MHz with external magnetic fields, the transition frequency of a transmon can be varied over many GHz. The transmons featured in this work have a tuning range of approximately $2.5$~GHz \textendash ~$10$~GHz, allowing them to be tuned into proximity with all of the modes in the photonic crystal.

% In addition to this high degree of frequency tunability, transmons also exhibit very large light-matter coupling strengths when fabricated inside microwave resonators. 

Because of the low transition frequencies of transmon qubits, they cannot be read out via fluorescence or photon absorption, as would be standard for an atomic qubit. Instead, standard practice is to exploit off-resonant coupling to a microwave resonator with frequency $\omega_r$ to perform readout \cite{Blais:revmodphys}. In the dispersive regime, where $|\omega_q - \omega_r| \gg g$, and approximating the transmon as a two-level system, the effective Hamiltonian takes the form:
\begin{equation}\label{eq:dispersiveJC}
    H_{\mathrm{dispersive}} = \hbar \left( \omega_r + \frac{g^2}{\Delta} \sigma_z \right) a^{\dag}a + \frac{\hbar}{2} \omega_q' \sigma_z, 
\end{equation}
where $\Delta  = \omega_q - \omega_r$ and $\omega_q' = \omega_q + g^2/\Delta$ is the Lamb-shifted qubit frequency.
Crucially, the term $g^2 \sigma_z a^\dagger a / \Delta$ constitutes a qubit-state dependent frequency shift of the resonator, known as the dispersive shift. 
%\textcolor{blue}{For a more detailed description of the dispersive shift that takes into account the higher energy levels of the transmon, see Ref.~\cite{Blais:revmodphys}.} 
For a more detailed description of corrections to the magnitude of the dispersive shift due to higher energy levels of the transmon, see Ref.~\cite{Blais:revmodphys}.
Due to the native separation of energy scales between the coherent coupling strength $g$ of a transmon and the native decay rates, transmons can easily reach the so-called strong dispersive regime in which the dispersive shift is comparable to or exceeds the resonator linewidth $\kappa$. In this regime, the amplitude and phase of a monitor tone impinging on the resonator are strongly affected by the state of the qubit, and, as a result, homodyne or heterodyne measurements of this microwave tone yield QND measurements of the qubit state \cite{Blais:revmodphys}.

Many geometries of resonator have been used in conjunction with transmons; however, in this work we concentrate only on the case of capacitive coupling to CPW resonators \cite{Goppl_2008}, because these resonators have the demonstrated capability to implement a large variety of unconventional band structures \cite{Kollar:2019hyperbolic,Kollar:2019linegraph}. 
These systems routinely display $g/2 \pi$'s in the range of $50$ \textendash ~$300$~MHz for single CPW resonators. 
%The qubit lifetimes typically exceed $10\ \mu$s, and single-cavity internal quality factors well above $200$k are routine, so the atomic and photonic decay rates $\gamma$ and $\kappa$ are on the scale of $2 \pi \times 10$--$2 \pi \times 100$~kHz, or smaller, easily three orders of magnitude smaller than $g$.
In the case of an array of CPW resonators, such as the photonic crystal studied in this paper, the natural photonic degrees of freedom are normal modes, rather than individual cavity modes. 
% A typical normal mode will be delocalized over the entire device, and have $g_{\mathrm{norm}} \sim g_0/\sqrt{N}$,\wcl{add why is this true?} where $g_0$ is the coupling between a qubit and a single resonator, and $N$ is the number of resonators in the array. 
% In an array of $N$ resonators, which each individually have a coupling $g_0$ to a qubit, a typically normal mode will be delocalized over the entire device, and have $g_{\mathrm{norm}} \sim g_0/\sqrt{N}$, due to the $N$-fold increase in the mode volume over that of a single isolated resonator. 
In an array of $N$ resonators, a typical normal mode will be delocalized over the entire device, and due to the $N$-fold increase in the mode volume over that of a single isolated resonator, the light-matter coupling will become $g_{\mathrm{norm}} \sim g_0/\sqrt{N}$, where $g_0$ is the coupling between the qubit and the particular resonator in which it is located.
Therefore, a CPW array with $\mathcal{O}(50-100)$ resonators can still have dispersive shifts which are an appreciable fraction of a MHz at a detuning of a GHz. Hence, the dispersive readout techniques which are normally used on transmon qubits in single-mode resonators can also be applied in resonator-array photonic crystal systems, using the normal modes.

\subsection{Coplanar Waveguide Lattices}\label{subsec:CPWlattices}

Since superconducting microwave resonators can be defined lithographically with low disorder and high quality factors \cite{Houck:earlylattice, Painter_Markov}, it is relatively straightforward to produce large resonator arrays with $50$ or more sites \cite{Underwood:imaging, Painter_Markov, Painter_SSH, Kollar:2019hyperbolic,Fitzpatrick_2017}, and there are many different resonator geometries to choose from when designing an array, each with different advantages or disadvantages. Lumped-element resonators, for example, are extremely compact \cite{Painter_Markov, Painter_SSH}, which is favorable for large-scale arrays. 

In this work, we concentrate solely on the case of CPW resonator arrays because they have a series of properties which are ideal for creating novel waveguide QED systems: first, they are known to be fabricable at large-scale with low disorder \cite{Houck:earlylattice, Underwood:imaging}; second, the frequency and effective hopping rate of a CPW resonator is insensitive to the shape of the resonator \cite{Kollar:2019hyperbolic}; and third, their one-dimensional nature and compatibility with three-way couplers leads to photonic band structures which host quadratic bands, Dirac cones, and flat bands \cite{Kollar:2019hyperbolic,Kollar:2019linegraph}.
% first, they are known to be fabricable at large-scale with low disorder \cite{Houck:earlylattice, Underwood:imaging}; second, their one-dimensional nature and compatibility with three-way couplers leads to photonic band structures which host quadratic bands, Dirac cones, and flat bands \cite{Kollar:2019hyperbolic,Kollar:2019linegraph}; and third, the frequency and effective hopping rate of a CPW resonator is insensitive to the shape of the resonator

Flat bands, which are highly unusual in naturally occurring band structures, are ubiquitous in CPW lattices because the individual lattice sites are one-dimensional line-like objects which connect at their end points, rather than point-like objects. As a result, the lattices produced correspond to a special class of graphs known as line graphs \cite{Kollar:2019linegraph, Kollar:2019hyperbolic}, which are generalizations of the kagome lattice. 
Every line graph originate from an underlying root graph, for example, the kagome lattice is the line graph of graphene, and all the properties of the line graph can be derived from the underlying root graph \cite{Kollar:2019linegraph}. 
%Line-graph lattices are generalizations of the kagome lattice that always have flat bands.
The mapping from the root band structure to the line-graph band structure for both symmetric and asymmetric on-site wavefunctions was worked out in Ref.~\cite{Kollar:2019linegraph}, and both cases are guaranteed to have flat bands.

The insensitivity of CPW resonator frequencies and hopping rates to resonator shape allows the design of CPW lattices to be extremely flexible, making it possible for CPW arrays to realize resonator packings which do not obey the conventional rules of two-dimensional crystallography, making it possible to realize lattices in effective curved spaces \cite{Kollar:2019hyperbolic} or giving the flexibility to fit a chosen lattice into a standard chip package. 
%For example, the lattice we study here, which will be discussed in detail in Sec.~\ref{sec:device}, is a quasi-one-dimensional lattice with all hopping coefficients equal and translation symmetry only in one direction, but it can be fabricated with resonators that are not equally spaced, to maximize packing efficiency, and with a switchback pattern to the unit cells, shown in Fig.~\ref{fig:device_image}, in order to fit the one-dimensional lattice on a square chip. 
% For example, the lattice we study here, which will be discussed in detail in Sec.~\ref{sec:device}, is a quasi-one-dimensional lattice with all hopping coefficients equal and translation symmetry only in one direction.
For example, a quasi-one-dimensional lattice with translation symmetry only in one direction, such as the one we study here (and discuss in detail in Sec.~\ref{sec:device}), would require a very long rectangular chip if made using CPW resonators of uniform size and shape.
However, as shown in Fig.~\ref{fig:device_image}, it can be fabricated on a square chip by using a switchback pattern of unit cells, and the individual unit cells can be made compact by making use of resonators of different shapes.
\begin{comment}
\ak{Roughly here was the transition to detailed stuff, which will now go to the appendices.}

\ak{what we need here now are declarations of the most important facts.}

\begin{itemize}
    \item super breif description of TB model, t proprtional to freq, FW t is negative, HW T is mixed

    \item super breif description of the sign of the hopping, with flat bands on bottom. point to previous refs for the details. 
    
    \item there are beyond TB corrections. We use first-order here. 
    
    \item CPW is 1D object. makes line graph lattice. band structure not symmetric about single-res energy, so sign of the hopping. Line graphs special. Generalized kagome lattices with flat bands.

    \item frequancy dependence of g.
    
\end{itemize}

\ak{paragraph below likely gets largely replaced, or becomes the intro paragraph to the appendices.}

\end{comment}

In the remainder of this section, we briefly summarize the theory underlying CPW resonator lattices.
A more detailed description can be found in Appendix~\ref{app:CPW_Lattices}.
An isolated CPW resonator of length $L$ bounded by capacitors has cosinusoidal standing-wave eigenmodes with antinodes at both ends of the resonator. The corresponding mode functions are
\begin{equation}\label{eqn:VoltageFunction_main}
\bar{\Phi}_{\mu}(x) \propto \cos\left(\frac{\mu \pi x}{L} \right)\!,
\end{equation}
with one end of the resonator taken to be at $x=0$ for simplicity.
The integer $\mu$ labels the different harmonics of a single resonator, with the lowest-frequency, fundamental mode corresponding to $\mu = 1$.
This antisymmetric fundamental mode, with opposite values of $\bar{\Phi}$ at each end of the resonator, satisfies $L = \lambda/2$ and is thus known as the half-wave mode.
The second-harmonic mode, corresponding to $\mu = 2$, is symmetric and occurs at twice the frequency of the half-wave mode.
It satisfies $L = \lambda$, and is known as the full-wave mode.
%occurs at twice the frequency of the half-wave mode, and, as it satisfies $L = \lambda$, is known as the full-wave mode.
A capacitively coupled transmon at the end of the resonator couples to an antinode of all modes $\mu$, and the strength of the qubit-mode coupling, $g_{\mu}$, scales as $\sqrt{\omega_{\mu}\omega_q}$, increasing with higher mode numbers (see Appendix~\ref{app_subsec:freqg} for details).

The resonators in the CPW lattice presented in this work are butt-coupled through 3-way coupling capacitors, previously developed in Refs.~\cite{Houck_Nature_QS,Houck:earlylattice,Underwood:imaging,Kollar:2019hyperbolic}, where the center pin of three resonators end in close proximity to one another.
The capacitance between neighboring resonators, $C_c$, introduces a coupling term to the Hamiltonian, and the resultant CPW lattice can be well described as an effective photonic tight-binding model of the form \cite{Koch:AnnPhysBerl,Koch:TRS_breaking}
% \begin{equation} \label{eqn:TBmodel_text}
%     H_\mu =  \sum_{n} \hbar \omega_\mu a_n^{\dagger} a_n - \sum_{\langle n, n' \rangle} \hbar t_{\mu}a_n^{\dagger}a_{n'}.
% \end{equation}
\begin{equation} \label{eqn:TBmodel_text}
    H_\mu =  \sum_{n} \hbar \omega_\mu a_n^{\dag} a_n - \sum_{\langle n, n' \rangle} \hbar t^{(\mu)}_{n,n'}a_n^{\dag}a_{n'}.
\end{equation}
Here, $\omega_{\mu}$ is the resonant frequency of a single resonator, $\langle n, n' \rangle$ indicates neighboring resonators that share a coupler, and $t^{(\mu)}_{n,n'}$ is the effective hopping between those resonators.
% While this form assumes that $\omega_{\mu}$ and $t_{\mu}$ are identical throughout the lattice, the more general case is derived in Ref.~\cite{Koch:AnnPhysBerl}.

% The strength of the hopping parameter for a given resonator harmonic increases with $\mu$ and is given by
% \begin{equation}\label{eqn:t_main}
%  t_\mu  \approx -\frac{1}{\pi}\omega_1 \omega_\mu C_c Z_0,
% \end{equation}
% where $Z_0$ is the characteristic impedance of the resonator.

The sign of the hopping in CPW lattices is generally unusual, with long-wavelength normal modes occurring at \emph{high} energy and short-wavelength modes occurring at \emph{low} energy \cite{Houck:earlylattice,Koch:AnnPhysBerl, Kollar:2019linegraph}, for example, the kagome-like flat-band induced by the line-graph structure is always at the bottom of the spectrum \cite{Kollar:2019linegraph}. 
One way to understand this is that in the standard black-box quantization circuit picture, the coupling capacitor is equivalent to a modulation in the kinetic energy or the effective mass, rather than a potential energy barrier (see Ref.~\cite{Koch:TRS_breaking}).

The sign of $t^{(\mu)}_{n,n'}$ depends on the sign of the mode function at the end of the resonators~\cite{Koch:AnnPhysBerl, Kollar:2019linegraph} (See Appendix~\ref{app_subsec:cpwarrays}). For the full-wave modes, which are symmetric, $t^{(2)}_{n,n'}$ is uniformly negative, which we denote by $t_2$.
For the half-wave modes, which are antisymmetric, $t^{(1)}_{n,n'}$ can have mixed sign. For analogy with the full-wave modes, we define an overall hopping parameter $t_1 = - \left \vert t^{(1)}_{n,n'} \right \vert$. The strength of the hopping parameter increases with $\mu$ and is given by 
\begin{equation}\label{eqn:t_main}
 t_\mu  \approx -\frac{1}{\pi}\omega_1 \omega_\mu C_c Z_0,
\end{equation}
where $Z_0$ is the characteristic impedance of the resonator.

Remarkably, Ref.~\cite{Kollar:2019linegraph} showed that for the symmetric three-way couplers utilized to make the CPW lattice presented in Section~\ref{sec:device}, the two resulting band structures formed by $H_1$ and $H_2$ both exhibit flat bands at $\omega_\mu - 2 |t_\mu|$, and the remaining bands are related by a band inversion.

% \textcolor{blue}{
% %While the strength of the hopping parameter increases with $\mu$ due to the dependence on $\omega_{\mu}$, 
% The sign of $t^{(\mu)}_{n,n'}$ for a given mode number is dependent on the parity of the resonator mode function.
% For symmetric mode functions, such as the full-wave modes,
% the sign of the hopping is uniformly negative.
% However, for antisymmetric mode functions, such as the half-wave modes, the sign of the hopping is mixed and can vary between different resonator pairs in the same lattice.}

% The strength of the hopping parameter for a given resonator harmonic increases with $\mu$ and is given by
% \begin{equation}\label{eqn:t_main}
%  t_\mu  \approx -\frac{1}{\pi}\omega_1 \omega_\mu C_c Z_0,
% \end{equation}
% where $Z_0$ is the characteristic impedance of the resonator.

% \textcolor{blue}{The sign of the hopping in CPW lattices is generally unusual, with long-wavelength normal modes occurring at \emph{high} energy and short-wavelength modes occurring at \emph{low} energy \cite{Houck:earlylattice,Koch:AnnPhysBerl, Kollar:2019linegraph}, for example, the kagome-like flat-band induced by the line-graph structure is always at the bottom of the spectrum \cite{Kollar:2019linegraph}. 
% One way to understand this is that in the standard black-box quantization circuit picture, the coupling capacitor is equivalent to a modulation in the kinetic energy or the effective mass, rather than a potential energy barrier (see Ref.~\cite{Koch:TRS_breaking}).}

For the values of $t_\mu/\omega_\mu$ achieved by the device presented here, corrections to the simple tight-binding description in Eq.~\ref{eqn:TBmodel_text} begin to become relevant.
A complete description of CPW lattices includes corrections that extend beyond the simple tight-binding model introduced in Eq.~\ref{eqn:TBmodel_text}.
In this work, we account for these corrections with a first-order perturbative approximation that the hopping rate is proportional to the \textit{normal-mode} frequency instead of the bare resonator frequency, which is described fully in Appendix~\ref{app_subsec:freqhopping}.
This correction compresses bands that appear below $\omega_{\mu}$ and stretches bands that appear above $\omega_{\mu}$.

\section{Device Design}\label{sec:device}

%\ak{Cite new photonic flat band int paper (Gonzalez Tudela). Maybe ehud altman Kagome?}Through the CPW resonator array on this device, we have engineered a diverse photonic band structure.
%\ko{Transition to CPW waveguide section before device design; 1D falls off exponentially with distance, Benedetto has sign changing flat band interaction, we have frustration and also flat bands. There exist many potential geometries...}
%\ak{mention other funkly lattices Cite new photonic flat band int paper (Gonzalez Tudela - sawtooth). Maybe ehud altman Kagome (cold atoms)?}
%On the other hand, the gapped flat bands in lattice considered here are generalized versions of the well-known Kagome-lattice flat band~\cite{Kollar:2019linegraph} and consist entirely of oscillatory states which flip sign between neighboring cites. The combination of the oscillatory photonic wavefunctions and the traingular plaquettes of the lattices can lead to a short-range frustrated magnetic interaction.
%\ak{trim som redundant stuff from here now that we mention line graphs above.}
The CPW resonator lattice platform can host a wide range of potential connectivities, with the associated band structures setting the geometric dependence of the qubit-qubit interactions. Unlike previous experiments \cite{Sundaresan} which looked only at simple 1D chains and quadratic band edges which give rise to exponentially localized qubit-qubit interactions, CPW lattices natively host quadratic, linear, and flat bands, and a much larger variety of interactions.
% The flat bands produced in these lattices are generalized versions of the well-known Kagome-lattice flat band which consist entirely of oscillatory states which flip sign between neighboring sites \cite{Kollar:2019linegraph}.
The line-graph flat bands produced in these lattices consist entirely of oscillatory states which flip sign between neighboring sites \cite{Kollar:2019linegraph}.
The combination of the oscillatory photonic wave functions and the triangular plaquettes of the lattice can lead to a short-range frustrated magnetic interaction.
Similar results have been shown theoretically for a 1D sawtooth lattice \cite{Benedetto2024}, but CPW lattices provide a route to extending this type of behavior to 2D and non-Euclidean cases.

To demonstrate the power of the CPW lattice platform, we implement a simple quasi-1D lattice with a wide variety of bands including both quadratic and linear bands, as well as gapped and ungapped flat bands.
The combination of non-trivial connectivity within the unit cell, which gives rise to this large variety of bands, with the favorable bulk-boundary ratio and side access found in a lattice with 1D translation symmetry makes this an ideal first test case for multimode QED systems which combine CPW lattices and transmon qubits.
In the following section we describe in detail the two band structures of our  target quasi-1D lattice, which arise from the half-wave and full-wave modes of the CPW resonator. We then show how this lattice can be realized physically and how transmon qubits are incorporated into the design without introducing large on-site disorder.

\subsection{Quasi 1D lattice with gapped flat bands}\label{subsec:peterchain}

\begin{figure}[th!]
\centering
		\includegraphics[width=0.45\textwidth]{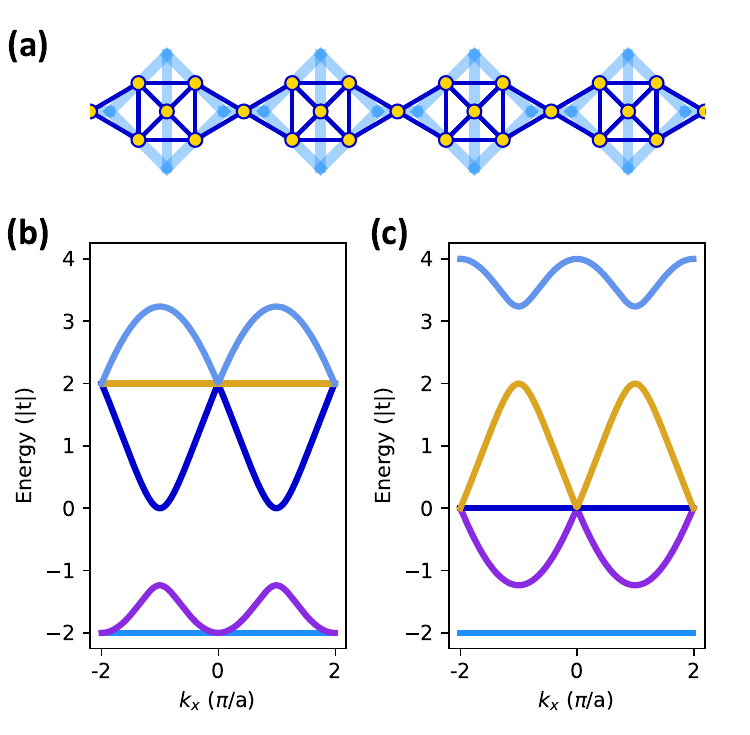}
	\caption{\textit{Target Lattice and its Band Structures.}
	(a) The target quasi-1D lattice overlaid on top of the resonator chain which generates it. The lattice sites are indicated in gold with dark blue lines indicating nearest-neighbor connections. The CPW resonator that produces a lattice site is indicated by a light blue line in the background. (b)-(c) Tight-binding calculation of the photonic bands for the half-wave and full-wave modes of the lattice, respectively, without frequency-dependent corrections. The energy of an uncoupled resonator mode is set to zero, and the bands are plotted in units of the hopping strength $|t_{\mu}|$. Each set of modes features two sets of flat bands (one doubly degenerate), 3 bands with conventional quadratic band edges, and one set of linear band crossings (a one-dimensional version of a Dirac cone).}
	\label{fig:targetlattice}
\end{figure}

%Multimodal waveguide QED systems allow for the exploration of interesting light-matter interactions, with the system’s dispersion relation establishing the type of interactions that can be observed. 
%For example, the spatial dependence of photon-mediated qubit-qubit interactions is set by the shape of the band mediating the interaction. 

Our quasi-1D lattice consists of rhombus-shaped unit cells arranged in a 1D chain.
Four unit cells of our target lattice are depicted in Fig.~\ref{fig:targetlattice}(a). All lattice sites have equal resonant frequencies (on-site energies), and all nearest-neighbor pairs have equal hopping strengths, up to experimental disorder \cite{Underwood:imaging}.
By utilizing resonators with different shapes and harnessing the flexible properties of the CPW resonator, this quasi-1D pattern can be produced on a square chip while ensuring that each resonator has the same frequency and preserving the 1D translation symmetry.

%As alluded to above, the primary motivation for choosing this target lattice is its interesting band structure. Experimentally, we only have access to the bands that form around the lowest two fundamental modes of our CPW resonators; the half-wave and full-wave modes. Since the sign of the hopping strength ($t$) in Eq. \ref{eq:TBmodel} is dependent on the parity of the normal mode (Eq. \ref{eq:CPWhopping}), the half-wave and full-wave modes produce different band structures. 

The band structure that arises from the coupled half-wave modes is shown in Fig.~\ref{fig:targetlattice}(b). There are two distinct sets of bands present; the higher set features a linear band crossing with a flat band in the middle while the lower set features a dispersive band with a quadratic band edge that touches two degenerate flat bands. The model of the band structure depicted here represents a diagonalization of the tight-binding Hamiltonian in Eq.~\ref{eqn:TBmodel_text} without taking into account the frequency-dependent hopping correction from Appendix~\ref{app_subsec:freqhopping}.

As the sign of the hopping parameters depends on the parity of the mode function (Eq.~\ref{eqn:VoltageFunction_main}), the full-wave modes of this lattice have their own separate band structure \cite{Kollar:2019linegraph}, depicted in Fig.~\ref{fig:targetlattice}(c). The highest band is an isolated dispersive band with a quadratic band edge and the middle set of bands has another linear band crossing around a flat band. Lastly, the lowest bands are two doubly-degenerate gapped flat bands. 
%While gapped flat bands are extremely rare in general lattices, they are common in CPW lattices as they arise from non-bipartite resonator connectivity
Such gapped flat bands are extremely rare in general lattices but common in CPW lattices, where they arise from non-bipartite resonator connectivity \cite{Kollar:2019hyperbolic,Kollar:2019linegraph}. 

Higher harmonics of the CPW, beyond the half-wave and full-wave modes, produce their own band structures, but these modes all have odd or even parities that match either the half-wave or full-wave mode functions. Thus, the only manner in which their resultant band structures differ is in the magnitude of their hopping parameters. Furthermore, the maximum frequency of the tunable transmons in our devices does not approach these higher energy bands.

\begin{comment}
This lattice was chosen due its band structure and simplicity. It is the simplest of its kind that goes beyond 1-D. The band structure consists of a gapped flat band, an ungapped flat band, a linear, and a quadratic band. Each band can facilitate interactions of different strengths and distances between qubits. 

-This lattice is cool an does crazy stuff.

-discuss specific HW and FW bandstructures we expect to see

The lattice features two different sets of photonic bands, one from the fundamental half-wave modes of the CPW resonators near $5$~GHz, and one from the second-harmonic full-wave modes at $10$~GHz. Due to the difference in the on-site wavefunction within each resonator, the two sets of modes exhibit different band structures, shown in Fig. \ref{bandsFig}. 

The chosen lattice exhibits many of the different types of bands possible in one dimension, making it an ideal choice for a first experimental prototype in which to explore qubit-mediated photon-photon interactions and their interplay with kinetic energy and dispersion. 
\end{comment}

% We have designed and fabricated a quasi one-dimensional lattice  of coplanar waveguide (CPW) resonators with nine unit cells, with each unit cell containing six resonators. The lattice sites are the resonators. To enable hopping between them, we coupled the resonators to one another capacitivey. The couplers connect three resonators at a time.

\subsection{Hardware Implementation}\label{subsec:hardwarelayout}
\begin{figure*}[t]
\centering
    \includegraphics[width=1\textwidth]{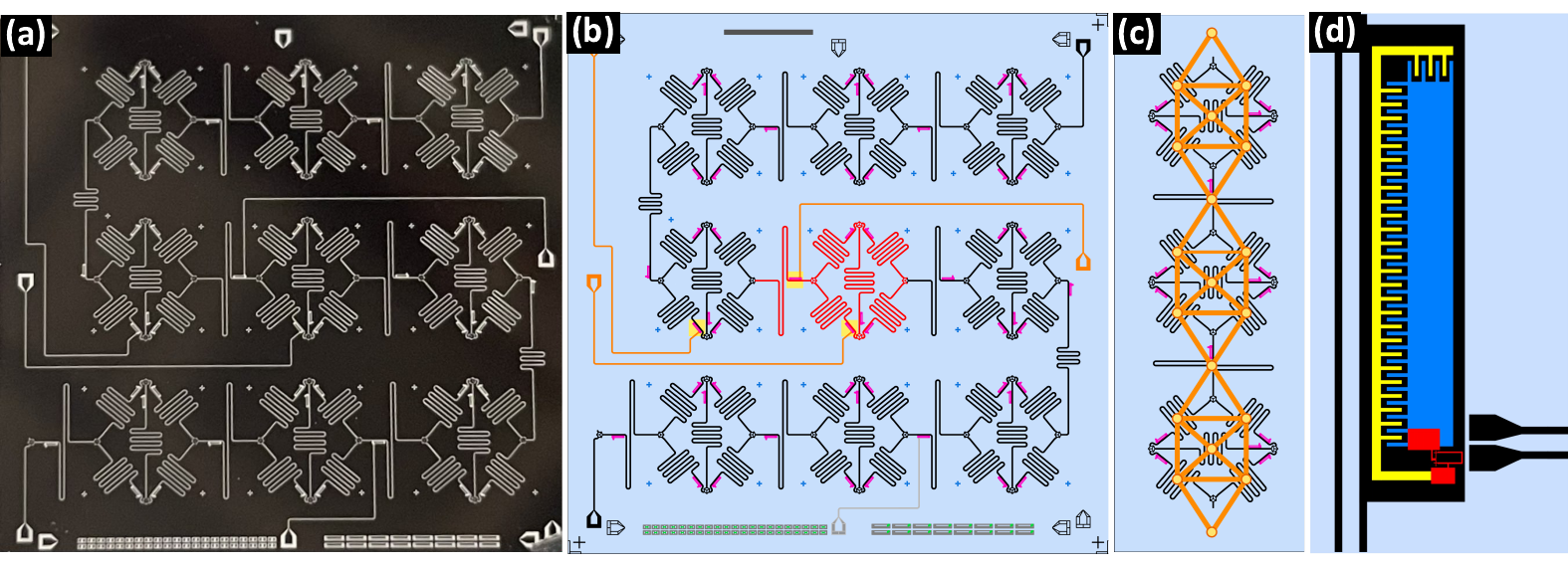}
	% \vspace{-0.6cm}
    \vspace{-0.2cm}
    \caption{\label{fig:device_image} 
    \textit{Device Design.} 
    (a) Photograph of the device after the photolithography stage. Dark regions are tantalum, and white regions are the underlying $25.4 \times 25.4 \ \mathrm{mm}^2$ sapphire substrate.
    %showing through the gaps in the Ta film which constitute the CPW resonators. 
    (b) CAD of the lattice device with color coding to illustrate different aspects of the design. The resonators forming the lattice are shown in black, with the exception of the central unit cell, which is highlighted in red. Input/output ports in the lower left and upper right of the device allow microwave transmission measurements. The locations of the three qubits are highlighted by the tan squares, and the associated on-chip flux bias lines, which allow independent frequency control of the qubits, are shown in orange. A fourth flux bias line (grey) is unused. The qubits are referred to as \sixteen, \five, and \eight ~from right to left. Transmon qubits, with the design shown in (d), are incorporated into the lattice using pockets opened in the side of the CPW shown in pink. To mitigate disorder, the capacitor paddles of a transmon are included in all 54 resonators, but the Josephson junctions needed to complete the transmons are only included at the three highlighted sites. (c) Effective tight-binding model describing photons in the device overlaid on the resonator network. (d) CAD design of the transmon qubit. The capacitor is digitated (yellow and dark blue regions), and the e-beam lithography pattern for the Josephson junctions is shown in red. The transmon is sandwiched between the center pin of the resonator just to its left (light blue) and its associated flux-bias line, the end of which can be seen in the lower right. 
    %\ak{as usual, my caption is too large.}
     % The qubits are referred to as \eight, \five, and \sixteen\text{ }from left to right. 
    }  
\end{figure*}

%In order to incorporate the quasi-1D design on a flat surface, 1$\times$1 in$^2$ tantalum/sapphire wafer is used to fabricate the design. Tantalum is used for superconducting metal for a ground plane and sapphire is used for insulating substrate. In Fig. \ref{fig:device_image}(a), there are 9 identical unit cells of the target lattice that are connected to each other. The starting and ending unit cells are connected to input and output port, forming a 1D chain. Figure \ref{fig:device_image}(b) shows the total CAD design of the device. A unit cell is highlighted in red color, clearly showing that it is composed of 6 resonators that is seen before in Fig. \ref{fig:targetlattice}(a). As a result, the total number of resonator on the device is 54. On-chip flux bias lines are colored with orange, routed to each transmon qubit from contact pad to tune a magnetic flux.

The device is fabricated on a $25.4\times 25.4 \ \mathrm{mm}^2$ sapphire wafer.
The CPW resonator lattice is then patterned on by using photolithography on a $200$~nm film of superconducting tantalum. 
%The CPW resonator lattice is written on a $25.4\times 25.4 \ \mathrm{mm}^2$ wafer of superconducting tantalum on an insulating substrate of sapphire using photolithography. 
The device after photolithography is shown in Fig.~\ref{fig:device_image}(a) and the CAD used during the fabrication process is displayed in Fig.~\ref{fig:device_image}(b). The device consists of nine unit cells of the quasi-1D lattice described in Section~\ref{subsec:peterchain}. The starting and ending unit cells are connected to input and output ports through which the device can be probed at microwave frequencies. (See Appendix~\ref{app:design} for details.) 
Each unit cell consists of six CPW resonators, highlighted in red in Fig.~\ref{fig:device_image}(b).
% A single unit cell is highlighted in red in Fig.~\ref{fig:device_image}(b); each cell consists of six CPW resonators, resulting in 54 total resonators on the chip. 
The resonators in the lattice are butt coupled using 3-way capacitors \cite{Houck:earlylattice,Kollar:2019hyperbolic}. To maintain uniform hopping, the coupling capacitors all have $120^\circ$ rotation symmetry, regardless of the orientation of the resonators, see Fig.~\ref{fig:res_test} for a more zoomed-in view. 
Fig.~\ref{fig:device_image}(c) shows an overlay of the lattice structure from Fig.~\ref{fig:targetlattice}(a) on top of a CAD image of the device to illustrate how the coupling scheme of the target lattice is physically realized by the hardware.
%Three-way symmetric capacitors couple resonators to their neighbors within the framework of the target lattice; when the capacitor only needs to couple two neighboring resonators, the third plate connects to the ground plane to introduce as little variation in the hopping between sites as possible. 
The lattice is laid out in an s-curve shape in order to accomodate a long 1D lattice on a square chip and uses four different resonator shapes to optimize packing density. This process of combining resonators with different form factors to produce a single lattice with equivalent sites, while in principle avoidable in this case, is intrinsically necessary for realizing fully general devices, such as non-Euclidean lattices, on chip.

Unlike previous realizations of CPW lattices \cite{Houck:earlylattice,Underwood:imaging,Kollar:2019hyperbolic}, this device incorporates transmon qubits into the resonator array. The design includes on-chip flux bias lines for up to four transmons capacitively coupled to lattice sites, three of which survived the fabrication process. The locations of the three transmons are highlighted in yellow in Fig.~\ref{fig:device_image}(b).
We label the qubits as \sixteen, \five, and \eight ~from right to left.
Each qubit has a corresponding on-chip flux bias line routed to it, colored orange in Fig.~\ref{fig:device_image}(b). DC current is applied via these lines to change the magnetic flux through the SQUID loops in the qubits, tuning their frequencies.
The CAD for these flux-tunable transmons is shown in Fig.~\ref{fig:device_image}(d). The tantalum capacitor paddles, depicted in dark blue and yellow, are defined using photolithography, along with the CPWs. 

The pocket in the ground plane needed to accommodate the transmon capacitor constitutes a significant perturbation to the host resonator, and would, if uncompensated, create a systematic on-site potential in sites which contain a qubit.
Therefore, the capacitor paddles of a transmon are included in \emph{all} 54 resonators.

In contrast, the SQUID loop (shown in red in Fig.~\ref{fig:device_image}(d)) which completes the transmon is included only in the highlighted sites. The SQUID and its constituent Al/Al$_2$O$_3$/Al Josephson junctions are written using electron-beam lithography and deposited using double-angle evaporation with in-situ oxidization.
More detailed information on the fabrication process is presented in Appendix~\ref{app:Fab}.

\section{Characterization of the Photonic Bands}\label{sec:bandcharacterization}

%In this section, we show how the modes of the resonator lattice can be located and characterized in situ using a combination of standard transmission spectroscopy and

In this section, we show how the modes of the resonator lattice can be located and characterized in situ using a combination of standard transmission spectroscopy and non-linear mode-mode spectroscopy.

% linear and non-linear spectroscopy. We combine standard transmission measurements and an extension of prior non-linear spectroscopy techniques \cite{Bosman_2017,Peugeot_2024, Blais:revmodphys}, which we term mode-mode spectroscopy, to identify all bands of the lattice.

% standard \textcolor{blue}{linear} transmission spectroscopy and \textcolor{blue}{non-linear spectroscopy}. 

% an extension of prior non-linear spectroscopy techniques \cite{Bosman_2017,Peugeot_2024, Blais:revmodphys}, which we term mode-mode spectroscopy.

% \textcolor{blue}{non-linear mode-mode spectroscopy, a new extension of Kerr spectroscopy~\cite{Bosman_2017,Peugeot_2024} and qubit two-tone spectroscopy~\cite{Blais:revmodphys}.}

%a \textcolor{blue}{generalized} implementation of two-tone spectroscopy, referred to as mode-mode spectroscopy. 

%In this section, we present data collected using two measurement techniques to probe the device's photonic modes: standard transmission and a novel implementation of two-tone spectroscopy referred to as mode-mode spectroscopy. These data sets are used to find and characterize the lattice's modes in situ and to identify modes which couple strongly to the qubits. 

\subsection{Transmission Measurements}\label{subsec:transmission}

The primary method for characterizing the photonic modes of the lattice is direct measurement of the transmission of microwave signals through the device versus frequency. In this method, the energy bands appear as frequency regions that are densely populated with high-quality-factor resonances. By comparing transmission spectroscopy of the bands with the structure of the target lattice, we confirm that the lattice maintains its mode structure in the presence of the incorporated qubits. Additionally, when the frequency of a qubit is tuned through the bands, the resonances that shift in response differentiate the true lattice modes from parasitic modes in the packaging that share  negligible coupling with the qubits. These characterization techniques are utilized for the half-wave modes (at \mbox{$\sim 5$}~GHz) and the full-wave modes (at \mbox{$\sim 10$}~GHz) to demonstrate that their band structures are consistent with Figs.~\ref{fig:targetlattice}(b) and (c) respectively.

%Less could, can, should, more definitive statements.
%Less usage of "we", use passive voice just about everywhere
%Less negativity, present the positives front and center. "This section felt like not wanting to get caught overpromising and overcompensating by underpromising"

\begin{figure*}[t]
\centering
    \includegraphics[width=1\textwidth]{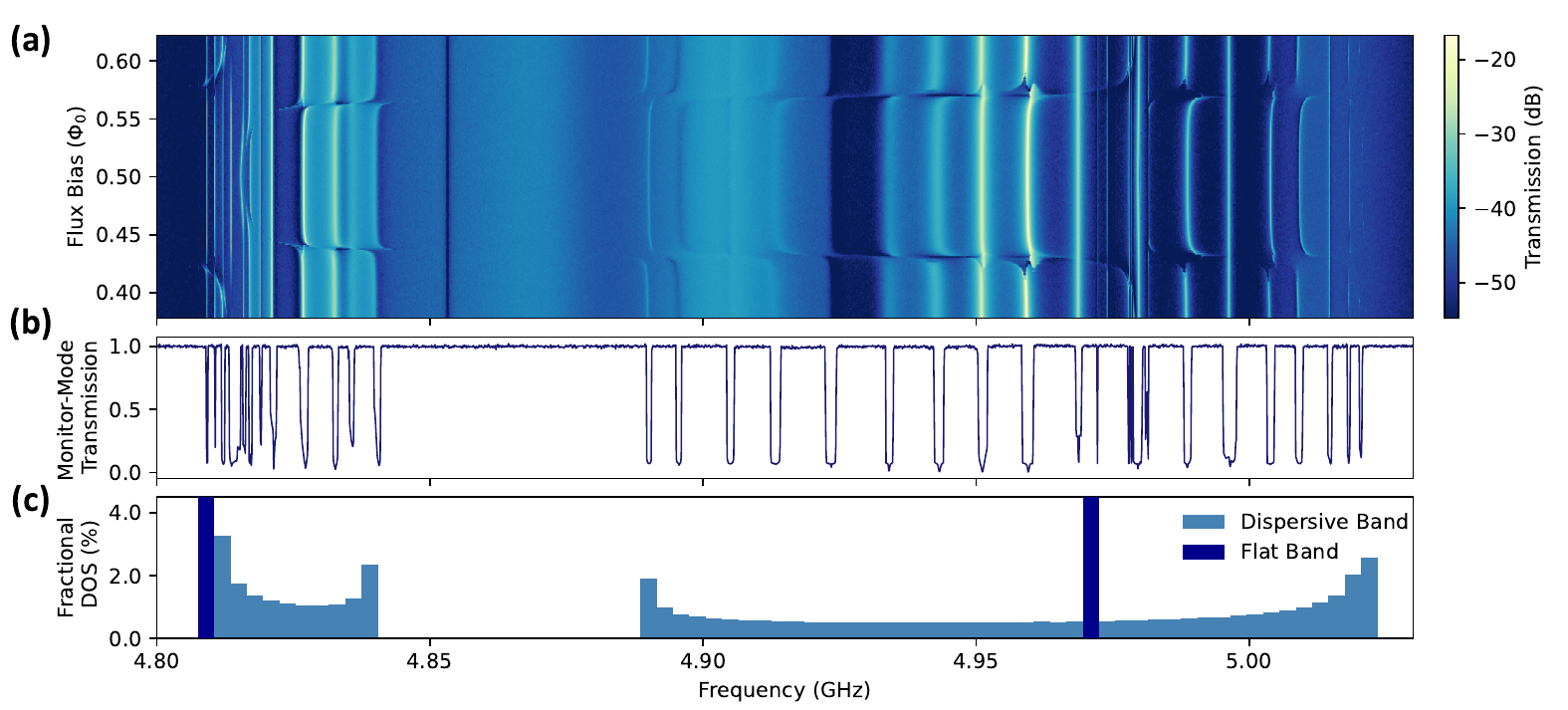}
    \vspace{-0.2cm}
	\caption{\label{fig:HWdata} 
    \textit{Half-Wave Characterization.}
    (a) Microwave transmission through the lattice as a function of flux applied to \eight, while the other two qubits are far detuned from the bands. The qubit \eight ~is tuned through the half-wave modes to its frequency minimum at half flux ($\phi_0 = 0.5)$, creating a series of avoided crossings. (b) Non-linear mode-mode spectroscopy of the half-wave modes, taken with all three qubits far detuned from the bands. Because this measurement relies on coupling to the qubits, it is insensitive to any weakly coupled parasitic modes in the device packaging. As a result, the band gap and two sets of bands are more clearly visible here than in the transmission data. (c) Expected density of states from the band structure of an infinite lattice and empirically determined values of the hopping parameter ($\vert t_1\vert/ 2\pi = 40$ MHz) and band center ($\omega_1/ 2\pi = 4.889$ GHz). Dispersive bands are shown in light blue, and the flat bands in dark blue. The full height of the flat-band peaks is well above the scale shown here. The two regions in which avoided crossings are visible in (a) correspond to the two frequency regions with non-zero density of states, and the mode-mode spectroscopy shown in (b) is in good agreement with the theoretical density of states.
    }  
\end{figure*}

A full transmission spectroscopy characterization of the half-wave bands is shown in Fig.~\ref{fig:HWdata}(a). 
Using the on-chip flux bias lines, the frequency of \eight ~is swept from above the half-wave modes to its frequency minimum at half flux and back again, while the other two qubits, \sixteen ~and \five, are held at a fixed frequency near $10$ GHz, far detuned from the half-wave modes. 
A transmission measurement is taken at each bias point to monitor how the features respond to the passage of the qubit. Both times that \eight ~crosses the half-wave bands, it leaves a trail of avoided crossings with the photonic modes while extraneous packaging modes are unaffected.
%(see for example the parasitic mode near $\sim 4.85$ GHz).
Comparing transmission spectroscopy of the half-wave modes with the predicted band structure from Fig.~\ref{fig:targetlattice}(b), the lower set of bands lies between \mbox{$\sim 4.81$}~GHz and \mbox{$\sim 4.84$}~GHz while the higher set of bands lies between \mbox{$\sim 4.89$}~GHz and \mbox{$\sim 5.04$}~GHz, with the two sets separated by a bandgap that is roughly $50$~MHz wide. 
%Low-power transmission scans for all three qubits are shown in Fig.~\ref{fig:HWzooms} in Appendix \ref{app:SuppTransmission}.
Equivalent transmission scans for all three qubits are shown in Fig.~\ref{fig:HWwide} in Appendix \ref{app:SuppTransmission}.
All three qubits show clear avoided crossings with modes in these two regions, although the coupling strengths to each individual lattice mode vary due to the different locations of the three qubits in the lattice.

\begin{figure*}[t]
\centering
    \includegraphics[width=1\textwidth]{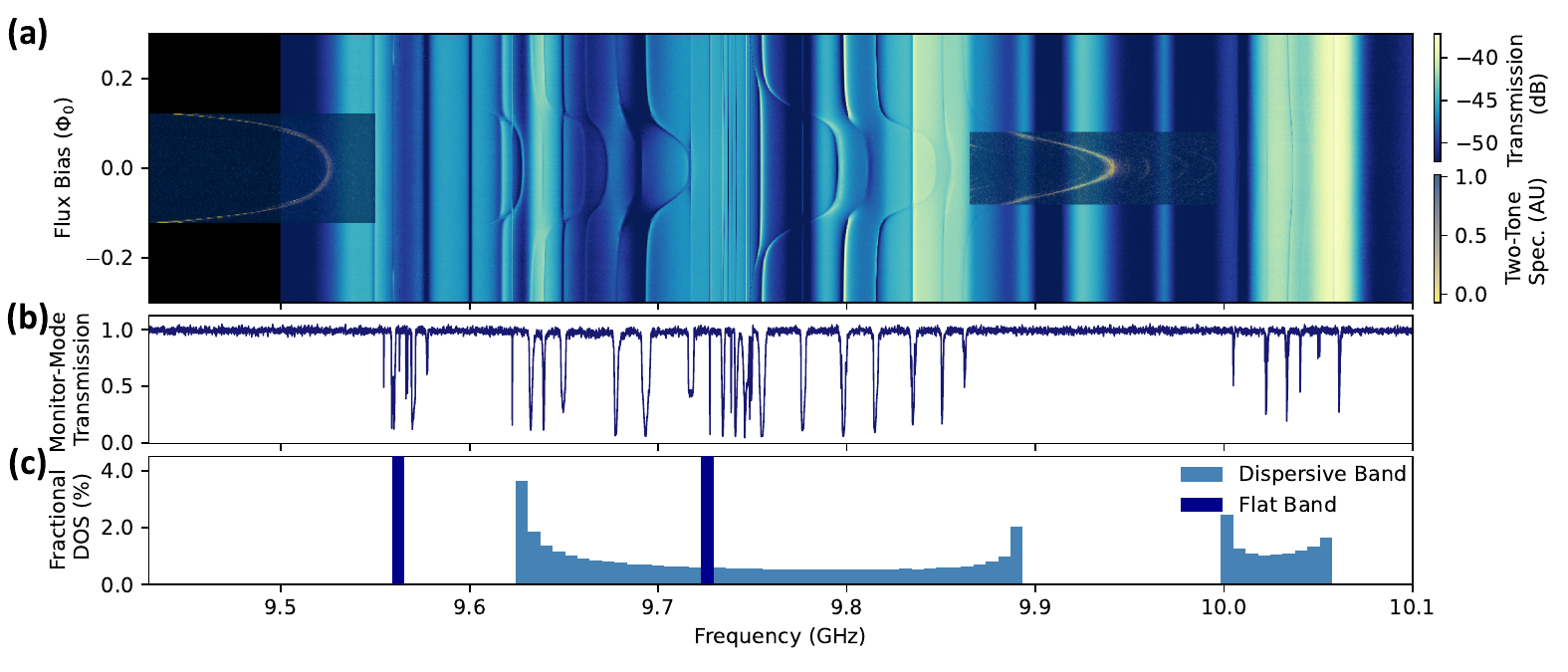}
    \vspace{-0.2cm}
	\caption{\label{fig:FWdata} 
    \textit{Full-Wave Characterization.} 
    Combined characterization data for the full-wave modes of the device. (a) Microwave transmission through the lattice as function of flux applied to \eight, while the other two qubits are held at fixed detuning. Near integer flux, the qubit \eight ~reaches its maximum frequency and tunes through the full-wave modes, creating a series of avoided crossings. In this frequency region, unwanted package modes are more significant than at the lower-frequency modes, resulting in a rolling background in transmission and causing many lattice modes to appear as Fano resonances. Also shown in (a), using a navy-gold color scale, is non-linear spectroscopy of the qubit transition itself. Two regions of this data are overlaid on top of the transmission, revealing qubit-like mid-gap states in which the qubit is heavily dressed by the the lattice modes. (b) Non-linear mode-mode spectroscopy of the full-wave modes, taken with all three qubits far detuned from the bands. The three sets of bands expected for this set of modes are easily distinguishable. Modes in the upper quadratic band and highly localized modes in the gapped flat bands that are difficult to observe in transmission are clearly visible in this mode-mode spectroscopy. (c) Expected density of states from the band structure of an infinite lattice using empirically determined values of the hopping parameter ($\vert t_2 \vert/ 2\pi = 82$~MHz) and band center ($\omega_2/ 2\pi = 9.726$ GHz). Dispersive bands are shown in light blue, and the flat bands in dark blue. The full height of the peaks due to the flat bands is well above the scale shown here. 
    }  
\end{figure*}

The transmission spectrum at the full-wave modes, shown in Fig.~\ref{fig:FWdata}(a), has two qualitative differences that distinguish it from the region near the half-wave modes.
First, the peak frequency of \eight ~at zero flux lies within the full-wave bands. 
Conveniently, this results in \eight ~being insensitive to flux while crossing the bands, a feature which facilitates observation of photonic bound states in the band gap through two-tone spectroscopy, which will be discussed in Section \ref{subsec:midgapspec}.
The second qualitative difference is the higher density of parasitic modes in the region.
Leakage through these modes produces a large broadband transmission signal, and for many of the full-wave modes, this leakage signal interferes with the transmission through the actual lattice, resulting in Fano-like resonances instead of the standard Lorentzian response. 
This is most clearly visible for the modes around $9.8$ GHz.
However, as the primary method for determining if a mode is part of the lattice relies on observing its response to the qubit, this change to the line shape does not hinder the process of categorizing the modes.

Comparing the photonic modes from the transmission spectroscopy in Fig.~\ref{fig:FWdata}(a) to the predicted band structure from Fig.~\ref{fig:targetlattice}(c), the middle three bands (purple, navy, and gold in Fig.~\ref{fig:targetlattice}(c)), which lie between $9.6$ GHz and $9.9$ GHz, are most clearly visible.
The remaining bands, the doubly-degenerate gapped flat band and the highest energy dispersive band, are difficult to locate in transmission due to their low coupling to the input and output ports of the device.
With the appropriate power settings and a high frequency resolution, it is possible to view some of the modes in these bands in transmission (see Appendix \ref{app:SuppTransmission}), but the bands can be much more cleanly visualized using a non-linear measurement technique that we term mode-mode spectroscopy, and which will be described in detail in Section~\ref{subsec:modespec}.
%\ko{I cut out reference to the specific locations of the gapped flat bands and highest energy band since you can't really tell from the transmission scan, so now have to point out in mode-mode spec section. Make sure to fill that in somewhere}

The well-resolved avoided crossings between \eight ~and the photonic energy bands in Fig.~\ref{fig:HWdata}(a) and Fig.~\ref{fig:FWdata}(a) indicate that the device has achieved strong qubit-mode coupling, despite the fact that the photonic modes are spatially extended throughout the lattice instead of confined to a single resonator. 
In conventional circuit-QED devices, qubit readout relies of two-tone spectroscopy to measure the response of a single-mode cavity to the qubit state \cite{Blais:revmodphys}. (See Appendix~\ref{app:twotone} for details.) One consequence of the strong qubit-normal-mode coupling observed here is that two-tone spectroscopy of \eight~can be carried out in this multi-mode environment using modes that display a large response to the qubit frequency.
%\textcolor{blue}{One consequence of this is that readout of \eight ~is feasible in this multi-mode environment by using normal modes of the lattice to perform two-tone spectroscopy, instead of conventional single-mode resonators \cite{Blais:revmodphys}.}
%
% One consequence of this is that two-tone spectroscopy of \eight ~is feasible in this multimode environment. 
% Viable modes for usage in two-tone spectroscopy can be identified from these transmission scans by selecting the individual modes that display the largest response to sweeping the qubit frequency. 
%
For example, from Fig.~\ref{fig:HWdata}(a), the mode near $4.99$ GHz exhibits a significant frequency shift in response to the proximity of \eight ~and functions well as a monitor mode.
% Equivalent transmission scans in which \sixteen ~and \five ~are swept through the half-wave and full-wave modes are shown in Appendix~\ref{app:SuppTransmission}. 
% These scans confirm strong coupling for those qubits as well, and are used to determine viable modes for performing two-tone spectroscopy on those qubits.
Viable monitor modes for performing two-tone spectroscopy on \sixteen ~and \five ~are determined from the transmission scans in Appendix~\ref{app:SuppTransmission}.

%The mode-mode spectroscopy, shown in (b), is strongly matching with this density of states plot within the error bar of hopping strength and band center.

% %%%%%%%%%%%%%%%%%%%%%%%%%%%%%%%%%%%%%%%%%
% \begin{figure*}[h]
% \centering
% 		\includegraphics[width=0.98\textwidth]{Figures/HW_and_FW.png}
% 	% \vspace{-0.6cm}
%     \vspace{-0.2cm}
% 	\caption{\label{fig:mainTransmission} 
%     \textbf{Transmission data.} 
%     \textbf{a} HW data. \textbf{b} FW data.  \ak{Now that I think about writing a caption, I should change this to a,b,c in one column and d,e,f in the other. then I can say a-c HW data for each of the 3 qubits. d-f for the FW.}  
%     } 
% \end{figure*}
% %%%%%%%%%%%%%%%%%%%%%%%%%%%%%%%%%%%%%%%%%

\subsection{Mode-Mode Spectroscopy}\label{subsec:modespec}

While transmission spectroscopy provides crucial insights about the photonic modes of the lattice, undercoupled modes that do not couple efficiently to both the input and output ports of the device are absent from the transmission spectrum.
%While transmission spectroscopy provides crucial insights about the photonic modes of the lattice, the requirement that a mode couples \textcolor{blue}{strongly} to both the input and output ports of the device means that many lattice modes (particularly highly localized ones) are absent from the transmission spectrum.
The most dramatic example of this absence are the doubly-degenerate gapped flat bands of the full-wave modes, where the modes are localized and nearly invisible in transmission. 
As a result, it is difficult to compare transmission measurements with quantitative models of the target lattice and extract experimental values for the hopping and band centers.
In this section, we present a non-linear spectroscopic measurement technique for probing the lattice modes, that we refer to as mode-mode spectroscopy, which is independent of the coupling between the output port and the mode, allowing for better comparison with theoretical predictions.

In essence, mode-mode spectroscopy is an implementation of continuous-wave two-tone spectroscopy, a standard circuit QED measurement technique for locating qubit transitions (see Appendix~\ref{app:twotone}), but modified to detect other lattice modes.
The technique is an extension of Kerr spectroscopy for superconducting circuits, as performed in Refs.~\cite{Peugeot_2024,Bosman_2017}, but using a stronger pump power to produce large, qualitative responses in order to detect the presence or absence of modes rather than using a weaker pump power to observe the smaller, more quantitative Kerr shift. In mode-mode spectroscopy, the transmission of a microwave signal through one photonic mode of the lattice (referred to as the monitor mode) is recorded while the frequency of a second pump tone is swept over the remainder of the modes. The transmons coupled to the resonator array add an element of non-linearity to the normal modes of the lattice. 
Therefore, when the drive tone is brought into resonance with one of the other photonic modes, the frequency of the monitor mode shifts. 
At low drive powers, there is a weak frequency response that originates from Kerr nonlinearity \cite{Elliott_2018,Peugeot_2024,Bosman_2017}. 
Mode-mode spectroscopy utilizes the high-power regime, in which a drive tone near resonance with a lattice mode "ionizes" the transmons by driving them from their ground states to one of their higher energy levels through a multiphoton process, resulting in a much larger response, maximizing the number of visible modes (see Appendix~\ref{app:ionization} for details) \cite{Martinis:ionization,Blais:ionization}.
%The mode-mode spectroscopy we present here is all taken in this transmon ionization regime (i.e., with a very strong drive tone) to maximize the number of visible modes.

Mode-mode spectroscopy scans of the half-wave and full-wave bands are displayed in Fig.~\ref{fig:HWdata}(b) and Fig.~\ref{fig:FWdata}(b) respectively.
Both data sets were collected using the mode at $\sim 4.96$~GHz as a monitor and with the frequencies of the three qubits tuned to between $6$ and $8$~GHz, though mode-mode spectroscopy measurements are generally insensitive to qubit frequency. 

Mode-mode spectroscopy has two key properties that provide an advantage over standard transmission. 
First, mode-mode spectroscopy is sensitive \emph{only} to modes which couple reasonably strongly to one of the qubits. 
Parasitic packaging modes common in transmission measurements typically have a much larger mode volume than the CPW modes of the lattice and therefore couple much more weakly to the transmons. This, combined with their relatively low quality factors (and thus peak resonant powers), means that package modes generally do not show up in mode-mode spectroscopy, making the technique incredibly background free. Consequently, the bandgaps in mode-mode spectroscopy are completely featureless, creating a stark contrast between band and bandgap.

The second beneficial property of mode-mode spectroscopy is that as long as a mode can be driven, i.e., if it couples to the input port, it can be measured.
%\textcolor{blue}{For a mode to be visible in transmission, there must be significant coupling to the input and output port, while mode-mode spectroscopy measurements only require that there is power circulating in the mode. As the normal modes of the CPW lattice have high quality factors, high intracavity power can be achieved even with a weak coupling to the input port.}
For a mode to be visible in transmission, there must be significant coupling to the input and output port. Mode-mode spectroscopy measurements, on the other hand, will work for undercoupled modes which can have significant intracavity power, but very low transmission.
%As the normal modes of the CPW lattice have high quality factors, high intracavity power can be achieved even with a weak coupling to the input port.}
Thus, some modes that are invisible in transmission, especially localized modes in the flat bands, can show up clearly in mode-mode spectroscopy.
This is most dramatically evident for the full-wave modes. In the transmission spectroscopy in Fig.~\ref{fig:FWdata}(a), neither the doubly-degenerate  gapped flat bands nor the band edges of the highest energy dispersive band can be accurately characterized in transmission due to the low input or output coupling of the modes. However, using mode-mode spectroscopy (Fig.~\ref{fig:FWdata}(b)), it is plainly apparent that the flat bands are located near $9.57$~GHz and the upper dispersive band lies between $10$~and $10.1$~GHz.

%\ko{Is this paragraph still ok? Some of these terms are only defined in freq-dependent hopping appendix now.}
%\ak{Stip out the reference to the frequency dependent hopping that are too specific}
Clear identification of bands and gaps from mode-mode spectroscopy makes it possible to compare the experimental characterization of the lattice modes with a theory model of the target lattice band structure to extract the mode frequencies, $\omega_{\mu}$, and the hopping strength, $t_{\mu}$, for the half-wave and full-wave modes ($\mu = 1,2$). 
Because the measurements are not momentum resolved, it is not possible to directly compare the experimental data to the theoretical band structures, shown in Fig.~\ref{fig:targetlattice}(b) and (c). Instead, we compare to the theoretical density of states to verify that the predicted bands and gaps correspond to the frequency ranges in which we observe the presence or absence of modes.
%\textcolor{blue}{As mode-mode spectroscopy probes the energy spectrum of the target lattice, it is most natural to compare these measurements with the density of states associated with the band structures of the target lattice (see Fig.~XYZ) rather than directly compare them with those band structures.}

For each resonator harmonic $\mu$, we compute the infinite-system normal mode frequencies numerically using a Bloch-wave ansatz and periodic boundary conditions, incorporating an approximation that the hopping rate is proportional to the normal-mode frequency (see Appendix~\ref{app_subsec:freqhopping} for details).
Fig.~\ref{fig:HWdata}(c) and Fig.~\ref{fig:FWdata}(c) depict the density of states calculated in this manner of the half-wave and full-wave bands. 
The parameters $\omega_{\mu}$ and $t_{\mu}$ are empirically determined by matching the band edges in the numerical density of states with the mode-mode spectroscopy data. For the half-wave bands, a hopping parameter of $\vert t_1 \vert/ 2\pi = 40$ MHz and on-site energy of $\omega_1/ 2\pi = 4.889$ GHz were extracted. The full-wave modes yield a hopping parameter of $\vert t_2 \vert/ 2\pi = 82$ MHz and on-site energy of $\omega_2/ 2\pi=9.726$ GHz.

%The experimental mode-mode spectroscopy data for both the half-wave and full-wave modes are compared to the tight-binding model with frequency-dependent hopping, as introduced in Section \ref{subsubsec:freqhopping}, to extract the fundamental mode frequencies, $\omega_{\mu}$, and the zeroth order hopping strength $t_{\mu}^{(0)}$, of our device. Fig. \ref{fig:HWdata}(c) and Fig. \ref{fig:FWdata}(c) depict the calculated density of states for the half-wave and full-wave bands of the target lattice. These are generated by numerically solving  

The observed agreement between the mode-mode spec data and the density of states model indicates that this device is a successful implementation of our target lattice. Consequently, it demonstrates that no significant faults were introduced during the design or fabrication process despite incorporating transmon qubits and laying out the resonator chain in the s-curve pattern required to fit the array onto a square chip.

\section{Qubit Characterization}\label{sec:qubitmeasurements}

In this section, we demonstrate the applicability of traditional circuit-QED qubit readout protocols in a multimode environment. We observe a qubit-photon bound state near resonance with the full-wave bands using two-tone spectroscopy, and using the same technique, we show an effective qubit-qubit interaction in the bandgap mediated by the lattice bands.

\subsection{Near Resonance}\label{subsec:midgapspec}

%Emphasize that this is first feature with qubit character; you can get two qubits interacting with each other in the bands but no lifetime - need to go into the bandgap (virtual photon exchange)
%Now we can say that we use these to get g and point to appendix.

Having verified the photonic modes of the device, we use conventional continuous-wave two-tone spectroscopy, as described in Appendix~\ref{app:twotone}, to measure the transition frequencies of the transmons embedded in the lattice and characterize their hybridization with the photonic modes.
As discussed in Section \ref{subsec:transmission}, monitor modes with relatively strong coupling to a qubit are preselected by identifying the modes which display the largest avoided crossings in transmission.

Since the qubits are highly flux-tunable, they can be brought into proximity with either the half-wave or the full-wave modes of the lattice.
Near resonance, strong hybridization occurs, resulting in a qubit-like photonic bound state in the bandgap.
Two-tone spectroscopy data of the midgap bound states formed by bringing \eight~near the full-wave modes of the lattice is shown in Fig.~\ref{fig:FWdata}(a), superimposed on transmission data in the same frequency and flux range. One bound state is observed below the full-wave modes and a second is observed emerging into the upper bandgap near $9.95$~GHz.

Direct observation of both of these bound states allows verification of two key aspects of the device.
First, while $g_0$, the coupling between a qubit and the mode of a single isolated CPW resonator, cannot be directly measured in the multimode lattice, the frequency splitting between the two bound states provides a strong constraint.
An approximate value of $g_0/ 2\pi = 165$~MHz for the full-wave modes (and $g_0/ 2\pi = 82.5$~MHz for the half-wave modes) is determined from comparison of the bound states and numerical simulation (see Appendix \ref{app:gfit} for details).

Second, while the theoretically-predicted gapped flat bands are not conclusively identifiable in transmission, the measured flux-dependence of the lower bound state cannot be explained solely by coupling to the dispersive bands of the lattice.
This observed flux-dependence requires either the presence of the flat band or a qubit-resonator coupling $g_0$ which is both much stronger than the device circuit parameters and inconsistent with the avoided crossings and bound state observed in the upper band gap (see Fig.~\ref{fig:gfig} in Appendix~\ref{app:gfit}).
Therefore, the bound state data indicates that the qubits are interacting with strongly-coupled flat-band modes in the predicted region.

\subsection{In the band gaps: photon-mediated interactions}\label{subsec:avoidedcrossings}

\begin{figure*}[t]
\centering
    \includegraphics[width=1\textwidth]{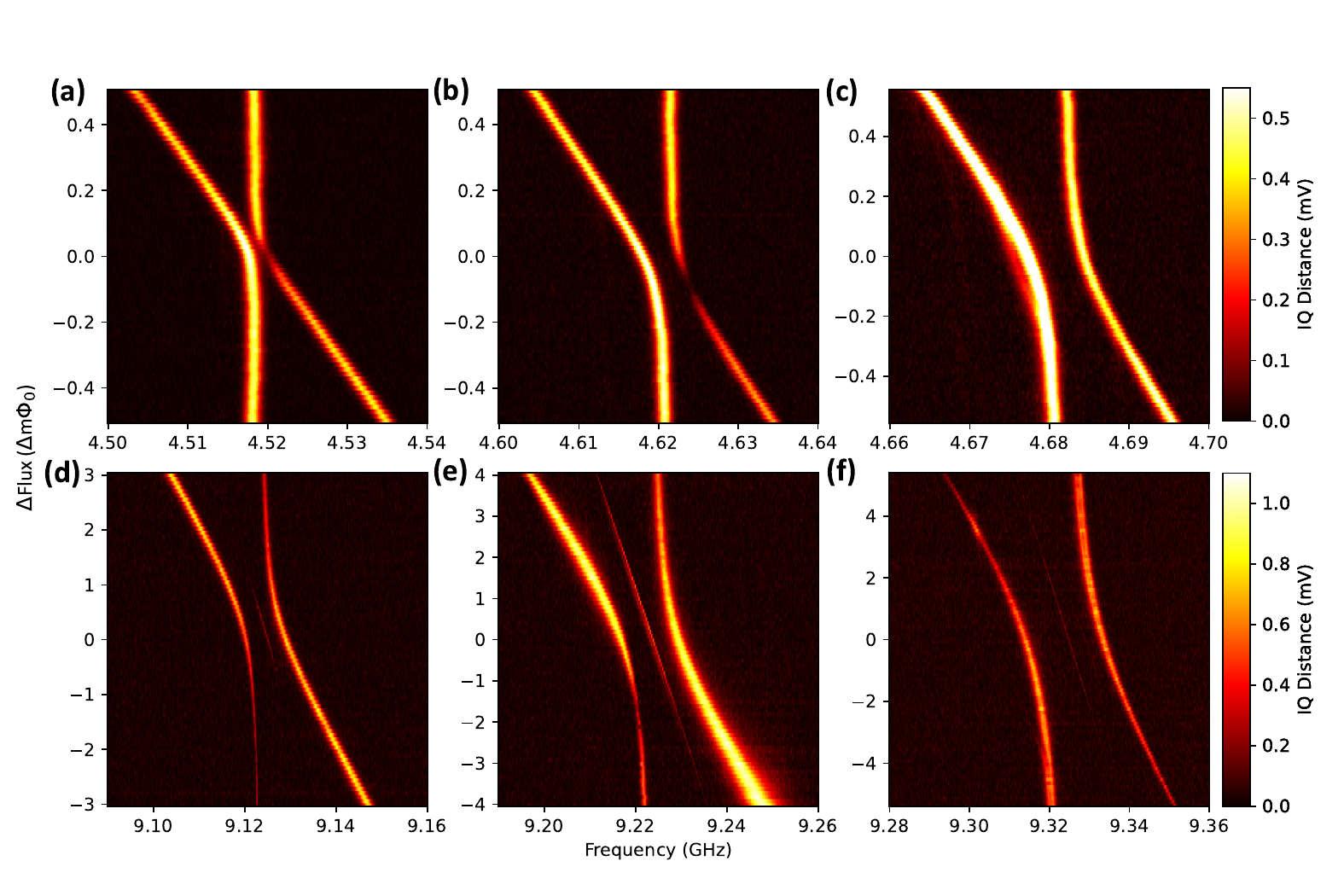}
    \vspace{-0.2cm}
    \caption{\label{fig:avoidedcrossing} 
    \textit{Photon-mediated avoided crossings.}
    (a)-(f) Avoided crossings between \sixteen ~and \five ~taken at six different frequencies. The flux applied to \five ~is held constant while \sixteen ~is swept through resonance. The monitor mode of each plot is selected so that both qubits can be read out with the same mode. The two qubits couple differently to the selected monitor mode and can produce signals in different quadratures of the measurement signal. In order to combine both quadratures into one signal, we find the vector-valued change in the monitor signal and plot the magnitude of this change, known as the IQ distance. For each plot, the y-axis denotes the offset of \sixteen ~from exact resonance with \five ~in units of milli-flux-quanta ($10^{-3}\Phi_0$). (a)-(c) show avoided crossings near the half-wave modes and (d)-(f) show avoided crossings near the full-wave modes. The strength of the interaction between the two qubit-photon bound states depends on their proximity to the closest band edge, with stronger interactions occurring as \sixteen ~and \five ~are brought closer to a set of bands. Also visible in (d)-(f), but most prominently in (e), is the two-photon transition from $\vert 00 \rangle$ to $\vert 11 \rangle$, which lies between the two main branches of the avoided crossing.
    }  
\end{figure*}

%The photonic modes of the device mediate interactions between qubits that share no direct coupling. The spatial dependence of these qubit-qubit interactions is set by the combination of the Wannier functions and the dispersion relation of the band mediating them. For example, the strength of interactions mediated by a quadratic band in a 1D chain falls off exponentially with distance.
%\ko{Transition to CPW waveguide section before device design; 1D falls off exponentially with distance, Benedetto has sign changing flat band interaction, we have frustration and also flat bands. There exist many potential geometries...}
%\ak{mention other funkly lattices Cite new photonic flat band int paper (Gonzalez Tudela - sawtooth). Maybe ehud altman Kagome (cold atoms)?}
%On the other hand, the gapped flat bands in lattice considered here are generalized versions of the well-known Kagome-lattice flat band~\cite{Kollar:2019linegraph} and consist entirely of oscillatory states which flip sign between neighboring cites. The combination of the oscillatory photonic wave functions and the traingular plaquettes of the lattices can lead to a short-range frustrated magnetic interaction.
Bound states near the band edges, such as those in Fig.~\ref{fig:FWdata}(a), have significant photon character, so while they can give rise to strong qubit-qubit interactions, the associated losses are unfavorable.
Moving deeper into the bandgap results in significantly lower losses and a qubit-qubit interaction which is dominantly mediated by virtual photons \cite{Calaj_2016}.
We exploit the frequency tunability of our transmons to measure qubit-qubit interactions in this regime, mediated by either set of modes in the lattice.

% whereas those mediated by a flat band... \ko{Need help with formal language to complete this thought, comments on previous draft indicate that we should tie back to Kagome lattice/frustration, not sure how to do concisely. Cite.}.
% \ak{Cite new photonic flat band int paper (Gonzalez Tudela). Maybe ehud altman Kagome?}Through the CPW resonator array on this device, we have engineered a diverse photonic band structure. Utilizing the frequency tunability of the transmons coupled to the lattice, we can explicitly tune pairs of qubits to hybridize with a specific set of bands and record the strength of the resulting interaction between two bound states.

To observe these interactions, one transmon is held at a fixed frequency near the bands of interest while the frequency of a second qubit is swept through resonance with it, using a feed-forward method to remove flux crosstalk (see Appendix~\ref{app:FluxCal} for details).
The resultant avoided crossing due to the virtual photon exchange is mapped out with two-tone spectroscopy. 
This procedure is carried out at three set detunings from the bottom of the half-wave bands (Fig.~\ref{fig:avoidedcrossing}(a)-(c)) and three detunings below the full-wave bands (Fig.~\ref{fig:avoidedcrossing}(d)-(f)), with \sixteen ~being tuned through resonance with \five ~in each of the six scans.
In each case, as the pair of qubits are brought closer to the mediating bands, the bound states grow less localized and their photonic wave functions overlap further causing the strength of the interaction to increase. 
This effect is observable through the size of the avoided crossings, which grow wider as the qubit pair approaches the bands.

The three crossings mediated by the full-wave modes in Fig.~\ref{fig:avoidedcrossing}(d)-(f) also have an observable multiphoton transition (most prominent in Fig.~\ref{fig:avoidedcrossing}(e)).
Each of the main branches of the crossing is a transition between $\vert 00 \rangle$ and a superposition of $\vert 01 \rangle$ and $\vert 10 \rangle$. 
The mid-crossing extra feature corresponds to an excitation from the $\vert 00 \rangle$ state directly to the $\vert 11 \rangle$ state using two drive photons.

In two-tone spectroscopy for a single qubit, measurement of a single quadrature of the monitor tone is typically sufficient to identify changes in the qubit state. 
However, for the avoided crossings depicted in Fig.~\ref{fig:avoidedcrossing}, where \sixteen ~and \five ~were measured simultaneously, each qubit has a distinct coupling to the selected monitor mode. 
Thus, a fixed excitation fraction for each qubit shifts the IQ components of the monitor tone differently, both in the size and direction of the IQ vector.
To maximize signal for both qubits, we measure the magnitude of the vector distance between the monitor tone and a reference signal in the IQ plane in order to capture both quadratures of the response in a single metric.

\section{Conclusion}\label{sec:conclusion}

In this work, we present the first CPW lattice device which incorporates both a lattice with non-trivial band structure and superconducting qubits. Our results show that low disorder photonic lattices, previously characterized in resonator-only devices \cite{Houck:earlylattice, Underwood:imaging,Kollar:2019hyperbolic}, survive the systematic introduction of qubits, even though transmon qubits are relatively large-area devices which can significantly perturb host resonators.

While our device realizes a quasi-1D lattice, which could in principle be realized using a CAD design with the same translation symmetry as the effective photonic tight-binding model, this would require an incredibly long rectangular chip. Instead, our device exploits the form-factor flexibility of CPW resonators to fit a non-square lattice on a square chip. This process of combining resonators with different form factors to produce a single lattice with equivalent sites, while in principle avoidable in this case, is intrinsically necessary for realizing non-Euclidean lattices on chip. Our results therefore indicate that hyperbolic CPW lattices can also be successfully combined with transmon qubits without introducing large amounts of systematic disorder.

We generalize measurement techniques developed for superconducting qubits in a single mode environment and show that they can be carried out in the highly multimode context of a photonic CPW lattice. These generalized techniques allow observation of photonic bound states in the lattice band gaps, due to hybridization of the qubits with the lattice modes, as well as virtual-photon-mediated interactions between qubits.

Furthermore, the generalized mode-mode spectroscopy technique presented here exploits any qubits in the lattice (and their associated Josephson nonlinearity) in order to implement fast and background-insensitive diagnostics of the photonic lattice, which were previously not possible. Previous resonator-only devices, such as those of Refs.~\cite{Houck:earlylattice, Kollar:2019hyperbolic} fundamentally lacked the ability to distinguish CPW lattice modes from any parasitic transmission channels in the device packaging. Scanning probe methods have previously been used to measure the normal-modes of a CPW lattice directly \cite{Underwood:imaging}. However, these methods are extremely time-consuming and require opening up the device package to leave the full top surface available for the scanning probe, which is likely highly detrimental to the lifetimes of any qubits in the lattice. By contrast, our non-linear spectroscopy method, while not a direct spatial probe, can be carried out in a fully-closed package and strongly distinguishes between active lattice modes and package leakage. 

In addition, mode-mode spectroscopy gives strong signatures as long as a mode is coupled to a qubit and can be driven. Hence, modes with weak input-output coupling and even highly-localized modes are strongly visible with this technique. Therefore, it provides a potential route to exploring the localization and delocalization of modes in a flat or hyperbolic band due to the influence of a local perturbing potential \cite{Galitski_Curvature}.

In general CPW lattices, the fundamental half-wave and second-harmonic full-wave modes exhibit distinct band structures, and only the second-harmonic band can exhibit gapped flat bands \cite{Kollar:2019linegraph}.
Therefore, realizing the full variety of spin-model connectivities made possible by combining photon-mediated qubit-qubit interactions and the graph-like flexibility of CPW lattices requires the ability to access both sets of modes and operate both qubits and measurements in two frequency ranges an octave apart. We demonstrate qubits which can span the required range and use conventional superconducting-qubit two-tone spectroscopy carried out via the normal modes of the lattice to observe virtual-photon-mediated interactions due to both sets of bands. 

In sum, the results presented here complete the toolkit developed by previous CPW lattice experiments and demonstrate the full experimental-design process and measurements needed to realize CPW lattices with superconducting qubits, both for Euclidean and non-Euclidean lattices.
Achieving graph-like flexible connectivity in CPW lattices is no longer limited to non-interacting resonator-only devices.
With the successful incorporation of qubits in this work, the door is now open to next-generation devices with larger numbers of qubits and 2D or hyperbolic \cite{bienias:Hyperbolic} spin-model connectivities, which can be used to implement driven-dissipative spin models with a large variety of connectivities.

% \ak{insert here?}
% This device, therefore, paves the way to driven-dissipative spin models with a large variety of connectivities. The door is now open to next-generation devices with larger numbers of qubits and 2D or hyperbolic \cite{bienias:Hyperbolic} spin-model connectivities. 

% \ak{Shown that qubits can be included in CPQ resonator lattice that has all of the complexity needed for hyperrbolic lattice. 3 way coupler. Variable resonator shapes. Localized and delocalize modes}

%Here we present an experimental platform suitable for implementing spin-spin interactions with a large variety of connectivities, spanning from simple 1D chains, to 2D, to positively-curved spherical spin models \cite{Houck_Nature_QS,bienias:Hyperbolic}, and even to hyperbolic spin models, by leveraging the flexible connectivity of microwave resonator arrays~\cite{Kollar:2019hyperbolic,Kollar:2019linegraph}. We demonstrate that transmon qubits can be integrated without compromising the versatility of the resonator array, as well as qubit-qubit interactions mediated by the bands of the array, completing the experimental toolkit needed to realize two-dimensional or hyperbolic spin models.

\begin{acknowledgments}
This work was supported by the National Science Foundation (QLCI grant OMA-2120757, PHY2047732, and PFC at JQI PHY-1430094), the Air Force Office of Scientific Research (Grant No. FA9550-21-1-0129), and the Sloan Foundation Research Fellow program, and the University of Maryland. MR received support from the LPS graduate fellowship and ARCS.

We thank experimental contributions from Theodore Gifford, Jeffrey Wack, Jake Bryon, and productive theoretical conversations with members of the Gorshkov group at UMD: Alexey Gorshkov, Yuxin Wang, Ali Fahimniya, and Alexandra Behne. We thank Trey Porto for insightful comments on the manuscript.
\end{acknowledgments}

\newpage
%\whitetext{blank page}
% \newpage

\appendix
%%%%%%%%%%%%%%%%%%%%%%%%%%%%%%%%%%%%%%%%%

\section{Coplanar Waveguide Lattice Theory}\label{app:CPW_Lattices}

The circuit theory of CPW lattices and the effective tight-binding description for photons in these lattices have been worked out in detail in Refs.~\cite{Koch:TRS_breaking, Jens_hopping,Koch:AnnPhysBerl,Kollar:2019linegraph}, so we will not provide a full treatment and set of derivations here. We instead summarize the key aspects of the CPW resonator modes and coupling structure necessary to show that any single resonator lattice configuration can give rise to two distinct band structures and to see the origins of beyond tight-binding corrections, which begin to be relevant for the hopping strengths present in our device. The key results outlined below closely follow the treatments of Refs.~\cite{Koch:AnnPhysBerl,Kollar:2019linegraph}.
%--- and something about onset of beyond TB model terms are large enough hoppings, which the device verges on.

\subsection{Single CPW Resonators}\label{app_subsec:singleCPW}

%\wcl{3/27 stopped here. Let's reread upto transmission}
Before examining the full complexity of a capacitively-coupled CPW resonator array, let us consider the mode structure of a single CPW resonator equipped with a coupling capacitor at each end. The description below follows the full derivations in Refs.~\cite{Koch:AnnPhysBerl, Koch:TRS_breaking}. The CPW itself is a free transmission line, consisting of a metallic center pin separated from two ground planes on either side by an insulating gap region, which supports traveling electromagnetic waves. For ease of circuit quantization, these modes are typically described in terms of the generalized flux operator $\Phi= \int{V \, dt} $ as a function of position along the center pin. Away from the boundary of the CPW, a mode at frequency $\omega_\mu$ is governed by the equation
\begin{equation}\label{eqn:CPWeigenmodes}
    \partial_x^2\Phi_{\mu}(x) = -\ell_{\ell} c_{\ell} \, \omega_{\mu}^2\Phi_{\mu}(x),
\end{equation}
where $\ell_{\ell}$ is the inductance per unit length of the CPW and $c_{\ell}$ is the capacitance per unit length. The coupling capacitance at the end of the CPW induces a boundary condition
\begin{equation}\label{eqn:CPWboundaryCond}
\pm \partial_x \Phi_{\mu}\vert_{x_{\pm}} = \ell_{\ell} C_c \, \omega_{\mu}^2\Phi_{\mu}\vert_{x_{\pm}},
\end{equation}
where $C_c$ denotes the coupling capacitance and $x_\pm$ denotes the locations of the two ends of the CPW resonator. 
This wave equation and boundary condition can be found from applying Kirchhoff's laws of circuits to a lumped-element description of a CPW. See e.g.~\cite{Koch:TRS_breaking}.

For an isolated CPW resonator, i.e. the limit $C_c \rightarrow 0$, Eq.~\ref{eqn:CPWboundaryCond} leads to the simple boundary condition $\partial_x \Phi_{\mu}\vert_{x_{\pm}} = 0$. Thus, unlike a particle in a box, or a conventional optical cavity, for which the mode functions go to zero at the edges, CPW resonator modes have \emph{antinodes} at each end. 
The frequencies of the eigenmodes of an isolated CPW of length $L$ are given by
\begin{equation}\label{eqn:normalfreqs}
    \bar{\omega}_{\mu} = \frac{\mu\pi}{L\sqrt{\ell_{\ell} c_{\ell}}},
\end{equation}
where $\mu$ enumerates the harmonics of the resonator, with the corresponding mode functions being
\begin{equation}\label{eqn:VoltageFunction}
\bar{\Phi}_{\mu}(x) \propto cos\left(\frac{\mu \pi x}{L} \right)\!,
\end{equation}
where we have taken one end of the resonator to be located at $x=0$ for simplicity.

The lowest-frequency, fundamental, mode corresponds to $\mu = 1$, and exists at a frequency for which $L = \lambda/2$. This mode is known as a half-wave mode and is antisymmetric, with opposite values of $\Phi$ at each end of the resonator. In analogy to solid-state systems, it can be thought of as being akin to a $p$-wave orbital. In contrast, the second-harmonic mode, with $\mu=2$, satisfies $L = \lambda$, and occurs at twice the frequency. 
%This full-wave mode has identical values of $\Phi$ at each end of the resonator, and can be thought of as an analog of a fully-symmetric $s$-wave orbital.
This full-wave mode has identical values of $\Phi$ at each end of the resonator, and can be thought of as an analog of a symmetric $s$-wave orbital \cite{Koch:AnnPhysBerl,Kollar:2019linegraph}. 
%\ak{I don't entirely like comparing this to a d-wave orbital, but any attempt to clarify that the internal structure doesn't matter, so it might as well be s-wave is combersome and it feels not helpful.}
% \ak{True. Orbital itself is mode d-wave like. But want to point people to s because the hopping is like that, an d-wave is complicated.}
%where $\mu =0$ corresponds to the lowest-frequency fundamental mode and $\mu = 1$ represents the second harmonic.

\subsection{CPW Arrays}\label{app_subsec:cpwarrays}

When multiple CPW resonators are fabricated on a chip, the ground planes between the resonators provide a significant amount of shielding, and coupling between resonators only becomes large if the center pins are brought into close proximity to one another with no intervening ground-plane region in between. In this work, we use a 3-way butt-coupled geometry, previously developed in Refs.~\cite{Houck_Nature_QS,Houck:earlylattice,Underwood:imaging,Kollar:2019hyperbolic}, in which the center pins of three resonators end in close proximity to one another. Since the coupling occurs at the end of the CPW, where generalized flux (voltage) is maximal and charge (current) vanishes, the resulting coupling is almost entirely capacitive, and the strength of the coupling can be tuned by varying the width of the center pin in the coupling region and the minimum separation between the center pins.

%\ak{insert comment about slight shifts when have a coupling cap and new freq $\omega'_\mu$ and slightly modified mode function}

The addition of this coupling capacitor has two effects: first, a slight shift in the resonant frequency $\bar{\omega}_\mu \rightarrow \omega_\mu$ due to the non-zero right-hand side in Eq.~\ref{eqn:CPWboundaryCond}, and second, introducing a coupling term between neighboring resonators.
The normal modes of a CPW array, which arise due to this capacitive coupling term, can be described as an effective photonic tight-binding model of the form
\begin{equation} \label{eqn:TBmodel}
    H_\mu =  \sum_{n} \hbar \omega_\mu a_n^{\dag} a_n - \sum_{\langle n, n' \rangle} \hbar t^{(\mu)}_{n,n'}a_n^{\dag}a_{n'},
\end{equation}
where $\omega_\mu$ is the resonant frequency of a single resonator, $\langle n, n' \rangle$ denotes nearest-neighbor pairs of resonators, and $t^{(\mu)}_{n,n'}$ denotes the effective hopping strength between any two resonators participating in a 3-way coupler. Here we consider only the case where all coupling capacitors are equal and all resonators have identical resonance frequencies. (The more general case is derived in Ref.~\cite{Koch:AnnPhysBerl}.)
A tight-binding Hamiltonian of this form is extremely general; 
however, the versions realized in CPW lattices have several universal and distinctive features.
First, because a coupling capacitor is equivalent to a modulation of the kinetic energy rather than a potential energy barrier, the energetics of the coupled modes are unusual, with long-wavelength normal modes occurring at \emph{high} energy and short-wavelength ones at \emph{low} energy \cite{Houck:earlylattice,Koch:AnnPhysBerl, Kollar:2019linegraph}. Second, the hopping strength is highly frequency dependent \cite{Koch:AnnPhysBerl, Kollar:2019linegraph}. Third, the half-wave and full-wave modes often give rise to different band structures, and in general two separate tight-binding models, with distinct $t^{(\mu)}_{n,n'}$, are required to describe both sets of normal modes in a CPW lattice \cite{Kollar:2019hyperbolic,Kollar:2019linegraph}.
%but the HW and FW modes can lead to \emph{distinct} band structures 

%\ak{If the coupling capacitance Cc leads to diferent resonator, you get hopping of the form.  End-to-end coupling might not be very understandable...}
All three of these properties arise from the fact that the coupling capacitance $C_c$ between two resonators which are end-to-end coupled gives rise to a hopping strength that can be approximated as
% All three of these properties arise from the fact that $t^{(\mu)}_{n,n'}$ due to single end-to-end coupling with capacitance $C_c$ can be approximated as
\begin{equation} \label{eqn:CPWhopping}
    t^{(\mu)}_{n,n'} = -\frac{1}{2} \omega_\mu C_c \tilde{\Phi}_{n,\mu} \tilde{\Phi}_{n'\!,\mu}
\end{equation}
where $\tilde{\Phi}_{n,\mu}$ denotes the value of the mode function $\Phi_\mu(x)$ in the $n^{\mathrm{th}}$ resonator evaluated at the end participating in the coupling, and  $\omega_\mu$ is the frequency of a single resonator.
%It immediately follows from Eq.~\ref{eqn:CPWhopping} that the hopping for the FW modes is twice as large 
For full-wave modes, the sign of $\Phi_2(x)$ is the same at both ends of the resonator, so the product $\tilde{\Phi}_{n,\mu} \tilde{\Phi}_{n',\mu}$ is always positive, leading to uniform \emph{negative} values of $t^{(2)}_{n,n'}$, which we denote by $t_2$. For half-wave modes, on the other hand, $\Phi_1(x)$ has opposite sign at opposite ends of the resonator. As a result, it is not always possible to choose a definition of $\Phi_1(x)$ on each resonator such that $\tilde{\Phi}_{n,1} \tilde{\Phi}_{n',1}$ is always positive. In these cases, $t^{(1)}_{n,n'} = \pm t_1$ will have constant magnitude but mixed signs.
%\ak{I haven't really brought up layout v effective lattices, so it's hard to state the bipartiteness condition for the two band structures to be the same. If feels slightly important, but also with a lot of machinery to even define what it is.}

% \subsection{Tight-Binding Hopping Parameters}\label{app:subsec:tbhopping}

% \textcolor{blue}{Following the arguments in Refs.~\cite{Koch:AnnPhysBerl, Koch:TRS_breaking}, the effective tight-binding hopping parameter for two capacitively coupled CPW resonators is given by:
% \begin{equation} \label{eqn:barebones_t}
%     t_{\mu} = -\frac{1}{2} \omega_\mu C_c \tilde{\Phi}_{\mu}^2,
% \end{equation}
% where $C_c$ is the coupling capacitance, and $\tilde{\Phi}_{\mu}$ denotes the value of the mode function on the two resonator ends participating in the coupling capacitor. 

To convert $t_\mu$ into a simpler form in terms of fundamental properties of the CPW and the harmonics of the resonator, we evaluate $\tilde{\Phi}_{\mu}$ using the normalization condition
$$c_\ell \int_0^L{\Phi_\mu^2(x)dx} = 1, $$where $c_\ell$ is the capacitance per unit length of the CPW and $L$ is the total length of the CPW resonator. Neglecting the small changes to the mode function due to the coupling capacitors, the mode function $\Phi(x) = A \cos (\mu \pi x  / L)$, and the normalization condition specifies $A = \sqrt{2 / c_\ell L}$, independent of $\mu$.
Substituting back in to Eq.~\ref{eqn:CPWhopping} yields:
\begin{equation}
t_\mu \approx -\frac{1}{2}\omega_\mu C_c A^2 =  -\omega_\mu C_c \frac{1}{c_\ell L}.
\end{equation}

However, the form above now depends on the
the capacitance per unit length $c_\ell$, which cannot be measured directly. A more convenient form can be obtained by eliminating $c_\ell$ and $L$ via their relationship to
the impedance $Z_0$ of the CPW transmission line and the fundamental frequency $\omega_1$ of the resonator \cite{pozar}: $L \sqrt{\ell_\ell c_\ell} = \pi / \omega_1$ and $Z_0 =  \sqrt{\ell_\ell/c_
\ell}$, where $\ell_\ell$ is the inductance per unit length of the CPW. Hence, $2/c_\ell L = 2 \omega_1 Z_0/\pi $, and 
\begin{equation}\label{eqn:t_z0}
t_\mu \approx -\frac{1}{\pi}\omega_1 \omega_\mu C_c Z_0 \approx -\frac{\mu}{\pi}\omega_1^2  C_c Z_0.
\end{equation}
Thus, the effective hopping rate scales linearly with the coupling capacitance and mode index $\mu$, but quadratically with the fundamental mode frequency.

\subsection{Frequency-Dependent Hopping}\label{app_subsec:freqhopping}
% The form of the hopping coefficient $t_\mu$ has frequency dependence both in $\omega_\mu$ and in the mode functions $\tilde{\Phi}_{n,\mu}$. 
% The exact mode functions for a CPW with a coupling capacitor are derived in Ref.~\cite{Koch:TRS_breaking,Koch:AnnPhysBerl}, but to lowest order, they can be approximated
% by the isolated-resonator mode functions in Eq.~\ref{eqn:VoltageFunction} in order to obtain a simple relation between $t_{\mu}$ and the fundamental properties of the CPW:
% % \begin{equation}\label{eqn:t_z0}
% % t_\mu \approx - 2 \pi \times 2 C_c Z_0 \frac{\omega_1}{2 \pi} \frac{\omega_\mu}{2 \pi},
% % \end{equation}
% \begin{equation}\label{eqn:t_z0}
% t_\mu \approx -\frac{1}{\pi}\omega_1 \omega_\mu C_c Z_0.
% \end{equation}
% where $Z_0$ is the impedance of the CPW. (See Appendix~\ref{app_subsec:HoppingParams} for details.) From this form, it follows that the hopping at the full-wave modes, $|t_2|$, is twice as large as the hopping at the half-wave modes, $|t_1|$.

%\ak{changed $\omega_\mu$ to $\omega_1$.}
As the ratio $t_\mu/\omega_1$ increases, additional corrections to the hopping rate begin to appear. The hopping strengths in the device presented here have $t_1/\omega_1 \approx 1\%$, and are just large enough that shifts in the band edges beyond the simple tight-binding approximation start to become visible. 
%\ak{try to reword the following} 
To simulate these effects, we use a first-order perturbative approximation that the effective hopping rate $t_{\mu}$ is proportional to the \emph{normal-mode} frequency, rather than the fixed bare frequency of a single isolated resonator.
% Motivated by the fact that the $\omega_\mu$ dependence in the hopping in Eq.~\ref{eqn:t_z0} arises from the frequency dependence of the impedance of the coupling capacitor, we take a self-consistent approximation that the effective hopping rate $t$ is proportional to the \emph{normal mode} frequency, rather than the fixed bare frequency of a single resonator.

In carrying out this approximation, the hopping parameter derived in Eq.~\ref{eqn:t_z0} is treated as the zeroth-order hopping rate, $t_{\mu}^{(0)}$. Diagonalizing the tight-binding Hamiltonian from Eq.~\ref{eqn:TBmodel} with $t_{\mu}^{(0)}$ gives the zeroth-order eigenenergies, which for normal mode $k$ in mode family $\mu$ are labelled $E_{\mu,k}^{(0)} = \hbar \omega_{\mu,k}^{(0)}$. These zeroth-order frequencies can be rewritten in terms of the hopping and the fundamental mode frequency:
\begin{equation}\label{eq:ZOEigen}
\omega_{\mu,k}^{(0)} = \omega_{\mu} + \epsilon_k t_{\mu}^{(0)},
\end{equation}
where $\epsilon_k$ is an implicitly defined scaling parameter. After replacing $\omega_\mu$ in Eq.~\ref{eqn:t_z0} with $\omega_{\mu,k}^{(0)}$ , the first-order approximation for the hopping is given by
\begin{equation}\label{FOHopping}
t_{\mu,k}^{(1)} =  \frac{t_{\mu}^{(0)}}{\omega_{\mu}} \omega_{\mu,k}^{(0)}.
\end{equation} 
Subsequently, an updated normal mode frequency, $\omega_{\mu,k}^{(1)}$, is found by inserting the effective hopping rate back into Eq. \ref{eq:ZOEigen}: 
% \begin{equation}\label{eq:FOEigen}
% \omega_{\mu,k}^{(1)} = \omega_{\mu} + \epsilon_k t_{\mu,k}^{(1)}.
% \end{equation} 
\begin{equation}\label{eq:FOEigen}
\omega_{\mu,k}^{(1)} = \omega_{\mu} + \epsilon_k t_{\mu,k}^{(1)} = \omega_{\mu} + \epsilon_k t_{\mu,k}^{(0)} + \frac{ \left[\epsilon_k t^{(0)}_{\mu,k}  \right]^2}{\omega_\mu}.
\end{equation}
In principle, this approach could be carried out to all orders to find a true self-consistent frequency. However, for the values of $t_\mu$ used here, higher-order corrections in $t_\mu/\omega_{\mu}$ are too small to be observed in the presence of experimental disorder in CPW lattices \cite{Houck:earlylattice}. We therefore consider only the first-order correction in Eq.~\ref{eq:FOEigen}.
This approximation, taking into account the frequency dependent impedance of the couplers, more accurately describes the normal modes of a CPW resonator array than the pure tight-binding model. 
In Section~\ref{subsec:modespec}, we use this model, applied to spectroscopic measurements of the bands of the lattice, in order to extract $\omega_{\mu}$ and $t_{\mu}^{(0)}$ for the half-wave and full-wave modes of our resonators.

Since transmission and mode-mode spectroscopy measurements are not momentum resolved, it is not possible to directly compare the perturbative approximation to the band structure with the experimental data. Instead, we compute the perturbed density of states in the limit of a large lattice with periodic boundary conditions and compare that to the frequency ranges in which we observe the presence or absence of modes. In particular, we optimize the theoretical parameters to match the locations of the edges of the band gaps.

%\ak{somewhere in the above we should mention the frequency dependence of g that goes with the freqency dependence of t. ratio g/t is roughly fixed. Maybe just a vey breif subsection 4 with 1 parapha.}

\subsection{Frequency-Dependent Coupling}\label{app_subsec:freqg}

Conventionally, the coupling strength $g$, as introduced in Eq. \ref{eqn:JC}, between a transmon and CPW resonator is written out in terms of the appropriate transmon charge matrix element and the RMS voltage in the resonator~\cite{Koch:transmon}.
It is uncommon to see the frequency dependence of the coupling explicitly included in this definition.
However, when working with higher harmonics of CPW lattices, it is crucial to understand how the coupling is affected when operating with different modes.
From Ref.~\cite{Koch:AnnPhysBerl}, the coupling strength for two resonators with frequencies $\omega_1$ and $\omega_2$ scales as $\sqrt{\omega_1 \omega_{2}}$, a fact which was used to derive the hopping between two identical CPWs in Eq. \ref{eqn:CPWhopping}.
Intuitively, this extends to a transmon coupled to a CPW resonator such that $g$ scales as $\sqrt{\omega_q \omega_{\mu}}$, since a transmon is a weakly anharmonic oscillator. Alternatively, this frequency dependence can be explicitly worked out from the standard charge matrix element and zero-point voltage form used in circuit QED \cite{Koch:transmon}.
%For a transmon coupled to a CPW resonator, the coupling strength $g$ for the qubit and resonator mode, as introduced in Eq. \ref{eqn:JC}, scales as $\sqrt{\omega_q \omega_{\mu}}$ \cite{Koch:transmon}.

Consequently, the coupling strength for a transmon on resonance with the full-wave mode of the resonator is about twice that of a transmon on resonance with the half-wave mode.
In the context of a resonator lattice, where the hopping scales linearly with $\omega_{\mu}$, the ratio $g/t$ remains roughly constant independent of the set of bands being observed.

\section{Supplementary Transmission Data}\label{app:SuppTransmission}

% \begin{figure*}[hb]
% \centering
%     \includegraphics[width=1\textwidth]{Figures/Fig. sup HW_Widescan.pdf}
%     \vspace{-0.2cm}
%     \caption{\label{fig:HWwide} 
%     \textit{High-power half-wave transmission.}
%     (a) Transmission data of the half-wave modes as a function of flux applied to \sixteen ~while the remaining two qubits are detuned from the range of interest. (b)-(c) Analogous data for flux applied to \five ~and \eight, respectively. At the relatively high input power used here, even very narrow weakly-coupled modes are clearly visible. However, avoided crossings in stronger modes (such as those near $4.96$~GHz) show glitches due to non-linear effects and transmon ionization. The sharp dip in midgap transmission near $4.85$~GHz matches the frequency of a non-topological edge state predicted by finite-size tight-binding calculations featuring the exact termination of our lattice.
%     }  
% \end{figure*}

%%%%%%%%%%%%%%%%%%%%%%%%%%%%%%%%%%%%%%%%%

%%%%%%%%%%%%%%%%%%%%%%%%%%%%%%%%%%%%%%%%%

%%%%%%%%%%%%%%%%%%%%%%%%%%%%%%%%%%%%%%%%%
\begin{figure*}[h!]
\centering
    \includegraphics[width=1\textwidth]{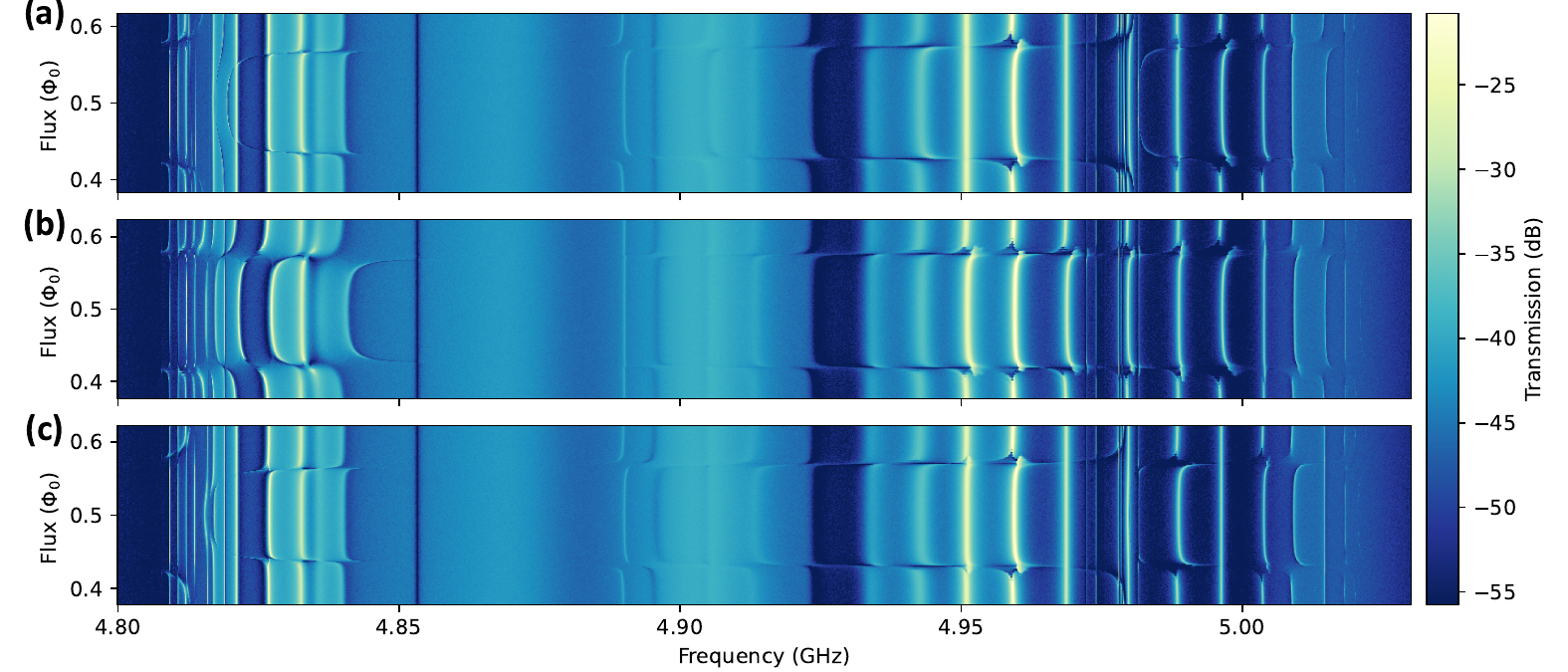}
    \vspace{-0.2cm}
    \caption{\label{fig:HWwide} 
    \textit{High-power half-wave transmission.}
    (a) Transmission data of the half-wave modes as a function of flux applied to \sixteen ~while the remaining two qubits are detuned from the range of interest. (b)-(c) Analogous data for flux applied to \five ~and \eight, respectively. At the relatively high input power used here, even very narrow weakly-coupled modes are clearly visible. However, avoided crossings in stronger modes (such as those near $4.96$~GHz) show glitches due to non-linear effects and transmon ionization. The sharp dip in midgap transmission near $4.85$~GHz matches the frequency of a non-topological edge state predicted by finite-size tight-binding calculations featuring the exact termination of our lattice.
    }  
\end{figure*}

\begin{figure*}[t]
\centering
    \includegraphics[width=1\textwidth]{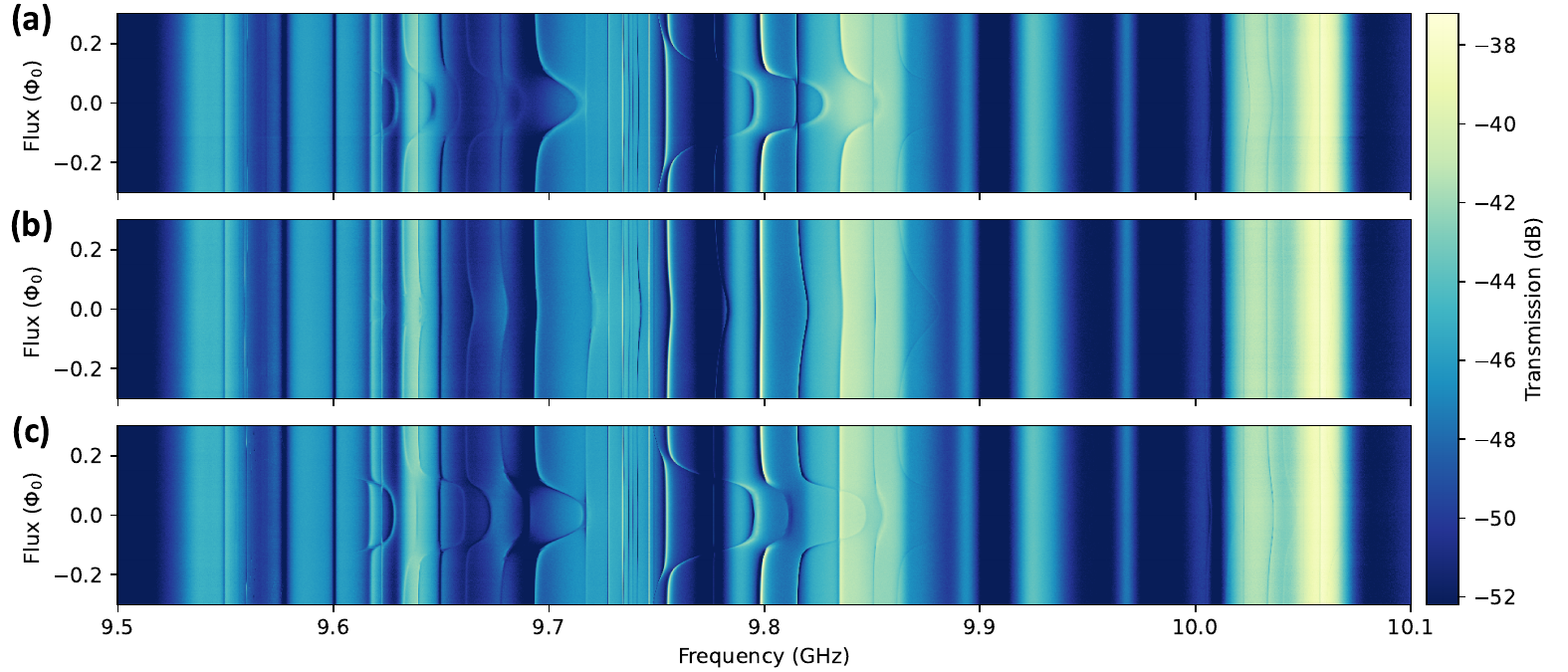}
    \vspace{-0.2cm}
    \caption{\label{fig:FWwide} 
    \textit{Full-wave transmission.} 
    (a) Transmission data of the full-wave modes as a function of flux applied to \sixteen ~while the remaining two qubits are detuned from the range of interest. (b)-(c) Analogous data for flux applied to \five ~and \eight, respectively. Our packaging contains a higher density of parasitic modes at these frequencies than at the half-wave modes. As a result, leakage transmission is a much more significant contribution in this region, leading to a broad rolling background and causing many modes to have Fano line shapes, rather than simple Lorentzians. The sharp dip in midgap transmission near $9.6$~GHz matches the frequency of a non-topological edge state predicted by finite-size tight-binding calculations featuring the exact termination of our lattice.
    } 
\end{figure*}
%%%%%%%%%%%%%%%%%%%%%%%%%%%%%%%%%%%%%%%%%

\begin{figure*}[ht!]
\centering
    \includegraphics[width=1\textwidth]{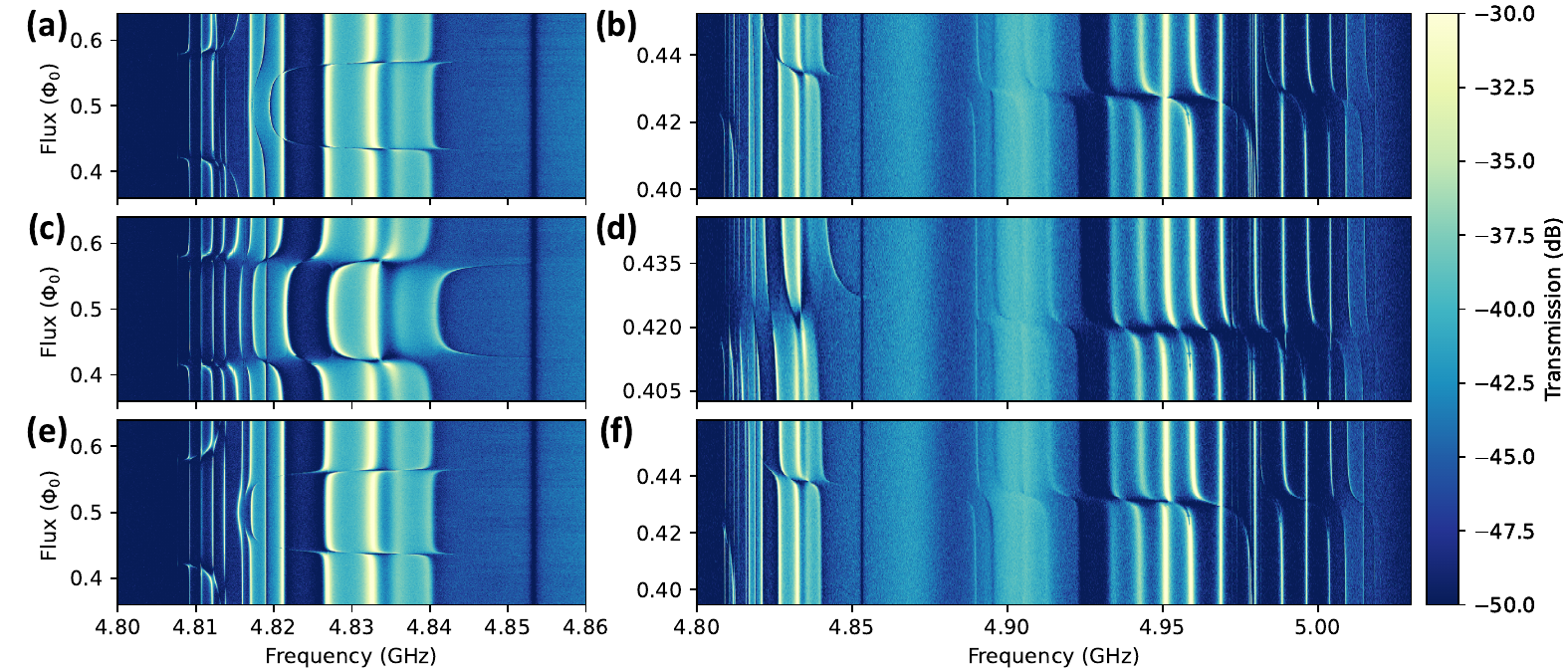}
    \vspace{-0.2cm}
    \caption{\label{fig:HWzooms} 
    \textit{Low-power half-wave transmission data.}
    (a)-(b) Transmission data of the half-wave modes as a function of flux applied to \sixteen ~while the remaining two qubits are detuned from the range of interest: (a) transmission magnitude near the degenerate flat bands and lowest-energy dispersive band and (b) transmission magnitude over the full set of half-wave modes. (c)-(d) Analogous data for flux applied to \five. (e)-(f) Analogous data for flux applied to \eight. 
    The mutiphoton-transition and transmon-ionization glitches seen in Fig.~\ref{fig:HWwide} are not present here due to the lower input power used ($5$~dB lower for the zoomed-in scans, $10$~dB lower for the wide scans), and the avoided crossings are smooth and continuous. 
    Due to the intra-unit-cell wavefunctions in the different bands and the inequivalent locations of the three qubits, \five ~does not couple to the half-wave flat bands, and as a result, the large avoided crossing visible for \sixteen~ and \eight~ in (a) and (e), is not present when tuning \five~ in (c). 
    Additionally, the mode at $4.95$~GHz is an example of a mode which has a node in one unit cell, and therefore does not couple to \sixteen ~in (b), but does couple to \eight ~in (f), which is located on the equivalent lattice site one unit cell over.
    % At the relatively low input power used here, glitches or multi-photon transitions (sometimes seen in Fig.~\ref{fig:HWwide}) are not present inside the avoided crossings.
    % avoided crossing do not show transmon-ionization-induced glitches, 
    % avoided crossings show fewer glitches as the drive tone is less likely to ionize the qubit than in Fig.~\ref{fig:HWwide}.
    % \ko{4/16: This caption needed work and there weren't comments left on it last pass-through, reworked it just now}
    }
\end{figure*}

\begin{figure*}[t!]
\centering
    \includegraphics[width=1\textwidth]{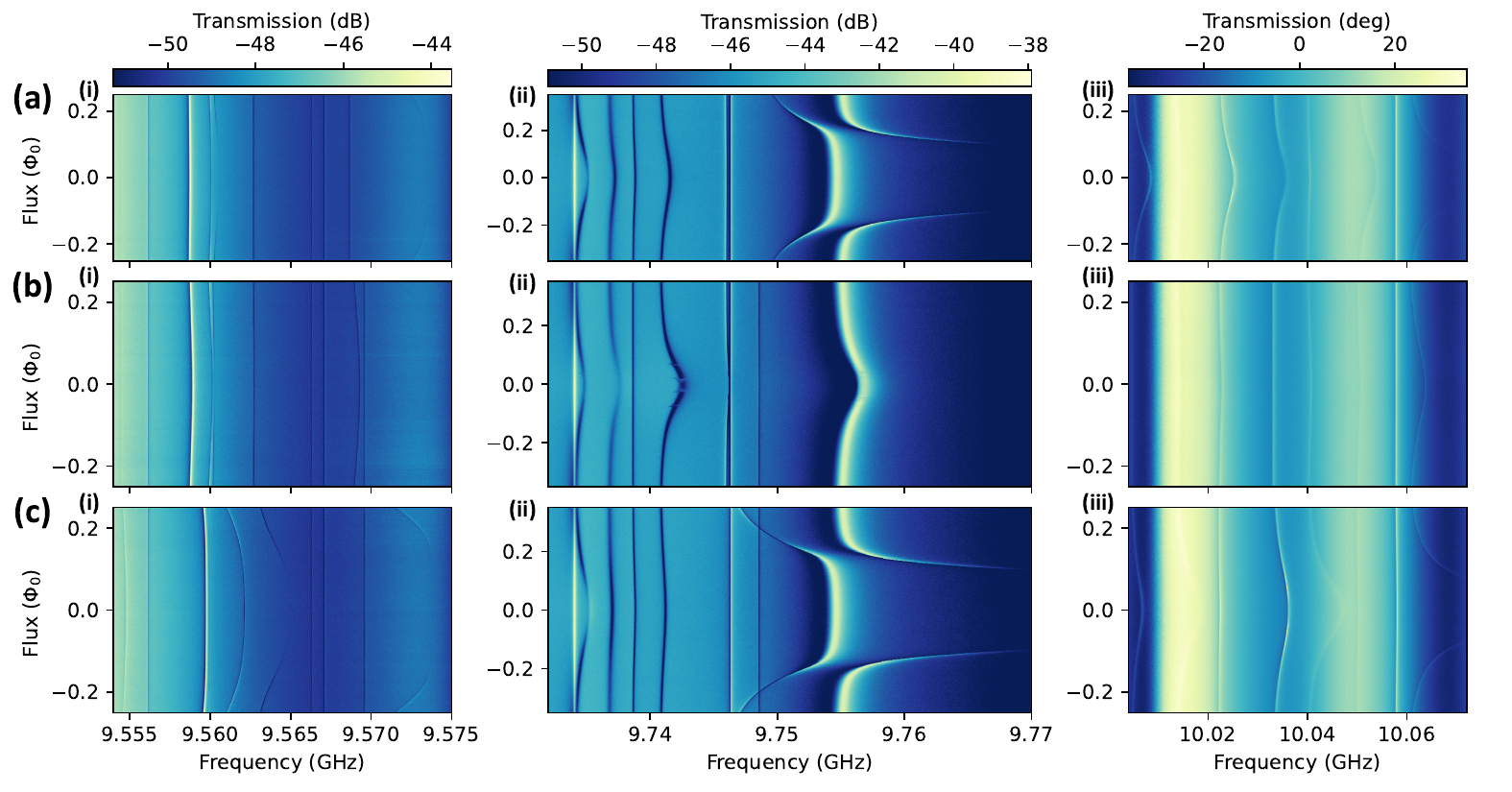}
    \vspace{-0.2cm}
    \caption{\label{fig:FWzooms} 
    \textit{Close-ups of band-edges and bands in full-wave transmission data.} 
    (a) Transmission data of the full-wave modes as a function of flux applied to \sixteen ~while the remaining two qubits are detuned from the range of interest. (b)-(c) Analogous data for flux applied to \five ~and \eight, respectively. The first column (i) shows the transmission magnitude near the gapped flat bands and column (ii) shows the region near the upper flat band. Column (iii) shows the highest dispersive band, but in this frequency range, the combination of large leakage and narrow linewidths leads to stronger signals in transmission phase than in transmission magnitude. Thus, only the phase is displayed for this band.
    Due to the intra-unit-cell wavefunctions in the different bands and the inequivalent locations of the three qubits, \five~ does not couple to the flat band near $9.75$~GHz, however, since \five~ does not ever fully enter the full-wave bands, the lack of a strong avoided crossing in (b.ii) is not a conclusive indicator of this effect. 
    On the other hand, \sixteen~ and \eight~ have similar maximum frequencies and (a) and (c) can be compared directly. The mode at $10.02$~GHz is an example of a mode which has a node in one unit cell, and therefore does not couple to \eight ~in (c.iii), but does couple to \sixteen ~in (a.iii), which is located on the equivalent lattice site one unit cell over.
    } 
\end{figure*}

%%%%%%%%%%%%%%%%%%%%%%%%%%%%%%%%%%%%%%%%%

%%%%%%%%%%%%%%%%%%%%%%%%%%%%%%%%%%%%%%%%%
% As mentioned above, large probe powers can lead to non-linear effects, so in transmission spectroscopy a tradeoff must be made between visibility of small features and clean avoided crossings.
% At high powers, modes are clearly visible with few averages and low-transmission modes can be detected with higher confidence.
% However, when a qubit is swept through resonance with a mode the drive tone has the potential to ionize the qubit (see Appendix \ref{app:ionization}) leading to glitches in the avoided crossings.
% At low powers transmon ionization is less likely but achieving equivalent SNR for small features is generally not possible, or extremely slow.
% The crossing between the half-wave mode at $\sim 4.96$ GHz and \eight ~is a perfect example of this effect of measurement-tone power.
% In the high-power scan (Fig.~\ref{fig:HWwide}(c)), the avoided crossing "breaks" when \eight ~is close to resonance with the mode, but in the scan with $10$~dB lower drive power (Fig.~\ref{fig:HWzooms}(f)), the full crossing is cleanly visible.

In Section \ref{subsec:transmission}, transmission scans of the half-wave and full-wave modes of our quasi-1D lattice device in which the frequency of \eight ~was tuned through resonance with the bands were presented. 
These scans were used to characterize qubit-mode coupling and identify suitable modes for usage in two-tone spectroscopy. Here we redisplay these scans alongside the equivalent transmission spectroscopy of the half-wave (Fig.~\ref{fig:HWwide}) and full-wave (Fig.~\ref{fig:FWwide}) bands for \sixteen ~and \five
, as well as zoomed-in scans of fine features at specific sets of bands in Figs.~\ref{fig:HWzooms} and~\ref{fig:FWzooms}.

% Additionally, we show zoomed-in transmission scans of fine features in the full-wave modes that are difficult to see in Fig.~\ref{fig:FWwide}, specifically of regions near the flat bands and highest dispersive band (Fig.~\ref{fig:FWzooms}).
% Some of the avoided crossings in Fig.~\ref{fig:HWwide} are obscured by the onset of nonlinear effects.
% Therefore, we also present more zoomed-in and lower power transmission scans of specific sets of bands within the half-wave modes (Fig.~\ref{fig:HWzooms}). 

Since large probe powers can lead to non-linear effects, a tradeoff must be made between visibility of small features and clean avoided crossings.
At high powers, modes are clearly visible with few averages and low-transmission modes can be detected with higher confidence.
However, when a qubit is swept through resonance with a mode the drive tone has the potential to ionize the qubit (see Appendix \ref{app:ionization}) leading to glitches in the avoided crossings.
At low powers transmon ionization is less likely but achieving equivalent SNR for small features is generally not possible, or extremely slow.
The crossing between the half-wave mode at $\sim 4.96$ GHz and \eight ~is a perfect example of this effect of measurement-tone power.
In a higher-power scan, such as Fig.~\ref{fig:HWwide}(c), the avoided crossing "breaks" when \eight ~is close to resonance with the mode, but in a scan taken with $10$~dB lower drive power (Fig.~\ref{fig:HWzooms}(f)), the full crossing is cleanly visible.

The three qubits in our device are located at and coupled to different resonators within the lattice, which lattice normal modes they individually couple to depends on the spatial distribution of the photonic wavefunctions.
By observing the size of avoided crossings between each normal mode and the three qubits, approximate locations of nodes and antinodes of the modes can be inferred.
For example, there are normal modes which have nodes at \sixteen~ and not \eight, even though these two qubits are located on equivalent sites within the unit cell: e.g., the half-wave mode at $4.95$~GHz (see Fig.~\ref{fig:HWzooms}(b) and (f)) and  the full-wave mode at $10.02$~GHz (see Fig.~\ref{fig:FWzooms}(a.iii) and (c.iii)).
However, quantitative comparison of the size of avoided crossings to tight-binding theory is challenging due to disorder and since qubit-normal-mode coupling strengths often exceed the normal-mode splitting. On the other hand, the structure of the unit cell leads to qualitative differences between how the three qubits interact with the different \emph{bands} of the lattice which are much more robust than the variations between individual modes.

%Due to the asymmetric on-site wavefunction for the half-wave modes, the corresponding tight-binding model does not fully respect the mirror symmetry of the connectivity, but the full-wave tight-binding model does. As a result, on-axis and off-axis qubits couple to the bands in qualitatively different ways. \ko{specify in HW modes?}
% -different bands have different weight on the inequivalent lattice sites within the unit cell.
% - most dramatic difference is the flat bands in the HW modes which do not touch Fm5.
% Each band has a different wave function within the unit cell

%, or ii) the full-wave modes, where all bands are either even or odd with respect to the mirror symmetry of the lattice, and the on-axis qubit only couples to even bands. This coupling difference in the full-wave modes is not directly visible in the transmission data since the peak frequency of the on-axis qubit is below the full-wave modes.
%The most dramatic differences in the qubit-mode couplings occurs within the flat bands of the device.

The connectivity of the lattice, shown in Figs.~\ref{fig:targetlattice}(a) and \ref{fig:device_image}(c), has a reflection symmetry about the center line.
The transmon \five ~is located in a resonator on this symmetry axis and is therefore referred to as on-axis.
The transmons \sixteen ~and \eight ~are located off-axis on the outer edges of the rhombus, at the same site in two neighboring unit cells.
Since the on-axis and off-axis qubits are located at inequivalent sites within the unit cell, they do not have the same coupling to each band. One dramatic difference occurs in the half-wave flat bands, the off-axis qubits, \sixteen ~and \eight, form very large avoided crossings with the doubly-degenerate lower flat bands in the half-wave modes between $4.81$ GHz and $4.82$ GHz, best visible in Fig.~\ref{fig:HWzooms}(a) and (e).
However, transmission spectroscopy of the same region while tuning \five ~(Fig.~\ref{fig:HWzooms}(c)) has no such feature.
This is consistent with the expected wave functions for modes in the degenerate flat bands, which are strongly present in off-axis resonators, and in the absence of disorder should have no amplitude in the connecting resonators between unit cells (including the resonator with the on-axis qubit \five). 
A similar phenomenon occurs around the upper flat band of the half-wave modes at $\sim 4.98$ GHz.
\sixteen ~and \eight ~form large avoided crossings around the location of the flat band (see Fig.~\ref{fig:HWzooms}(b) and (f)) while nothing of the sort is visible for \five ~(Fig.~\ref{fig:HWzooms}(d)).

% As mentioned above, large probe powers can lead to non-linear effects, so in transmission spectroscopy a tradeoff must be made between visibility of small features and clean avoided crossings.
% At high powers, modes are clearly visible with few averages and low-transmission modes can be detected with higher confidence.
% However, when a qubit is swept through resonance with a mode the drive tone has the potential to ionize the qubit (see Appendix \ref{app:ionization}) leading to glitches in the avoided crossings.
% At low powers transmon ionization is less likely but achieving equivalent SNR for small features is generally not possible, or extremely slow.
% The crossing between the half-wave mode at $\sim 4.96$ GHz and \eight ~is a perfect example of this effect of measurement-tone power.
% In the high-power scan (Fig.~\ref{fig:HWwide}(c)), the avoided crossing "breaks" when \eight ~is close to resonance with the mode, but in the scan with $10$~dB lower drive power (Fig.~\ref{fig:HWzooms}(f)), the full crossing is cleanly visible.

In the full-wave modes, the photonic tight binding model has uniform signs of the hopping parameters and respects the mirror symmetry of the connectivity. As a result, each band has fixed parity and, in the absence of disorder, only even bands couple to the on-axis qubit \five. However, the visible qualitative differences between the transmission spectroscopy for the off-axis qubits in Fig.~\ref{fig:FWwide}(a) and (c) and the on-axis qubit in Fig.~\ref{fig:FWwide}(b) ~originate not from symmetry constraints but from the fact that the peak frequency attained by \five ~at zero flux lies below the full-wave bands. \sixteen ~and \eight, on the other hand, enter deep into these bands.
The response of the highest dispersive band and modes near the gapped flat bands are not clearly visible in Fig.~\ref{fig:FWwide}. We therefore present zoomed-in transmission scans near the two flat bands and near the highest dispersive band for each of the three qubits in Fig.~\ref{fig:FWzooms}.
%Fig.~\ref{fig:FWzooms} shows zoomed-in transmission scans near the two flat bands and near the highest dispersive band for each of the three qubits.
%\ko{fine features hard to see in Fig. 7 near flat bands and highest dispersive band, easier to see in Fig...}

% %paragraph below to be replaced.
% In the full-wave modes, even though each band has fixed parity with respect to the mirror symmetry of the lattice, there are visible qualitative differences between the transmission spectroscopy for the off-axis qubits in Fig.~\ref{fig:FWwide}(a) \& (c) and the on-axis qubit in Fig.~\ref{fig:FWwide}(b) which primarily arise from the off-axis transmons having higher peak frequencies. \ko{reworded}
% The significantly more dramatic frequency shifts and avoided crossings observed for \sixteen ~and \eight ~originate not from symmetry constraints but from the fact that the peak frequency attained by \five ~at zero flux lies below thee full-wave bands. \sixteen ~and \eight, on the other hand, enter deep into these bands.

Numerical simulations of the photonic bound states, shown in Fig.~\ref{fig:gfig} and discussed in Appendix \ref{app:gfit}, indicate that \eight, and by similarity \sixteen, attain peak frequencies of roughly $9.85$\textendash $9.9$~GHz, which are inside the full-wave band structure.
As a result, modes in the highest dispersive band do not generally show full avoided crossings, merely deflections.
Modes in this upper dispersive band are particularly difficult to observe directly in transmission, due in part to the broad parasitic modes situated within the band.
For these modes, the phase response of the transmitted signal shown in Fig.~\ref{fig:FWzooms}(a)-(c)(iii) is stronger than the amplitude response that is typically used to characterize the other photonic modes.

\section{Two-Tone Spectroscopy}\label{app:twotone}

The dispersive readout mechanism described in Section~\ref{subsec:transmon_and_readout} can be used as a spectroscopy technique for locating qubit transitions by combining two drive tones: one to excite the qubit and one to monitor the qubit state. In the case of conventional single-mode devices, the fixed-frequency monitor tone is applied on-resonance with a resonator, or a normal mode in the case of the CPW lattice device studied here, whereas the drive/excitation tone is swept in frequency. 
This technique, known simply as two-tone spectroscopy, is standard in the field \cite{Blais:revmodphys}, and is conceptually simple in the idealized case in which the drive is known to implement a single $\pi$ pulse and state readout using the monitor tone is done in a time-resolved fashion. 
The base temperature of the dilution refrigerator naturally prepares the qubit in its ground state, and if the drive tone is resonant with the qubit transition, then the qubit will be excited,
causing the homodyne signal at the monitor frequency to switch to a different amplitude and phase. 

Pulsed drives and readout of this form require significant calibration and have low duty cycle due to the need to wait for the qubit state to reset between attempts, and as a result pulsed two-tone spectroscopy is generally slow. The continuous-wave version of two-tone spectroscopy is conceptually more subtle, but considerably faster and less sensitive to experimental parameters, so we employ this version to observe qubit transitions and photonic bound states in our CPW lattice device as a function of flux.

In particular, we use a fully continuous-wave protocol in which both the drive and monitor tones are applied continuously. In this case, if the drive tone is resonant with the qubit and strong compared to the qubit dissipation rate, $\gamma$, the qubit will be driven to the maximally mixed state. The measured microwave signal at the monitor frequency is thus an average of the values that would be obtained for a qubit purely in the $\ket{g}$ state and that of a qubit in the $\ket{e}$ state. The maximum achievable signal with continuous-wave two-tone spectroscopy is thus half of what is achievable in pulsed operation. However, this slight loss in signal can be vastly outweighed by the much greater robustness of the technique. For example, pulsed two-tone spectroscopy is very sensitive to the strength of the applied drive, and if the drive is too strong, it can apply an accidental $n \times 2 \pi$ pulse, and the qubit transition will no longer yield any signal. In continuous-wave all reasonably strong drives will saturate the qubit to the maximally mixed state. This drive-power insensitivity is particularly advantageous for looking at the photonic bound states. Because these are hybrid polariton states whose qubit-like character changes rapidly with detuning near the band edges, the effective Rabi rate induced by the drive can likewise change rapidly with frequency.
The robustness of mixed-state qubit measurements could instead be achieved with a pulsed setup by applying drive pulses long enough to saturate the qubit.
However, this method takes significantly longer than continuous-wave due to the reduced duty cycle of pulsed measurements, while still only achieving half the maximum achievable signal reached by coherent pulsed measurements, thus not offering a significant advantage over other methods.

Additionally, while dispersive readout is normally carried out only in the true dispersive regime with $|\Delta| \gg g$, where it is a QND state readout mechanism, two-tone spectroscopy can be used to find polariton/photonic bound state frequencies even for much smaller detunings. 
In the nondispersive regime the microwave signal does not provide a faithful measurement of the qubit state, but the monitor signal is independent of the presence of the secondary drive tone,
unless that drive tone excites a nonlinear qubit-like transition. Therefore, the \emph{presence} of a nonlinear response between the drive and monitor tones indicates the location of a qubit-like mode, even if the type of response is not a clean indicator of the qubit state. We exploit this effect in Section~\ref{subsec:midgapspec} to measure photonic bound states within $g_0$ of the band edge.

\section{General Device Design}\label{app:design}

\begin{figure*}[t!]
\centering
    \includegraphics[width=1\textwidth]{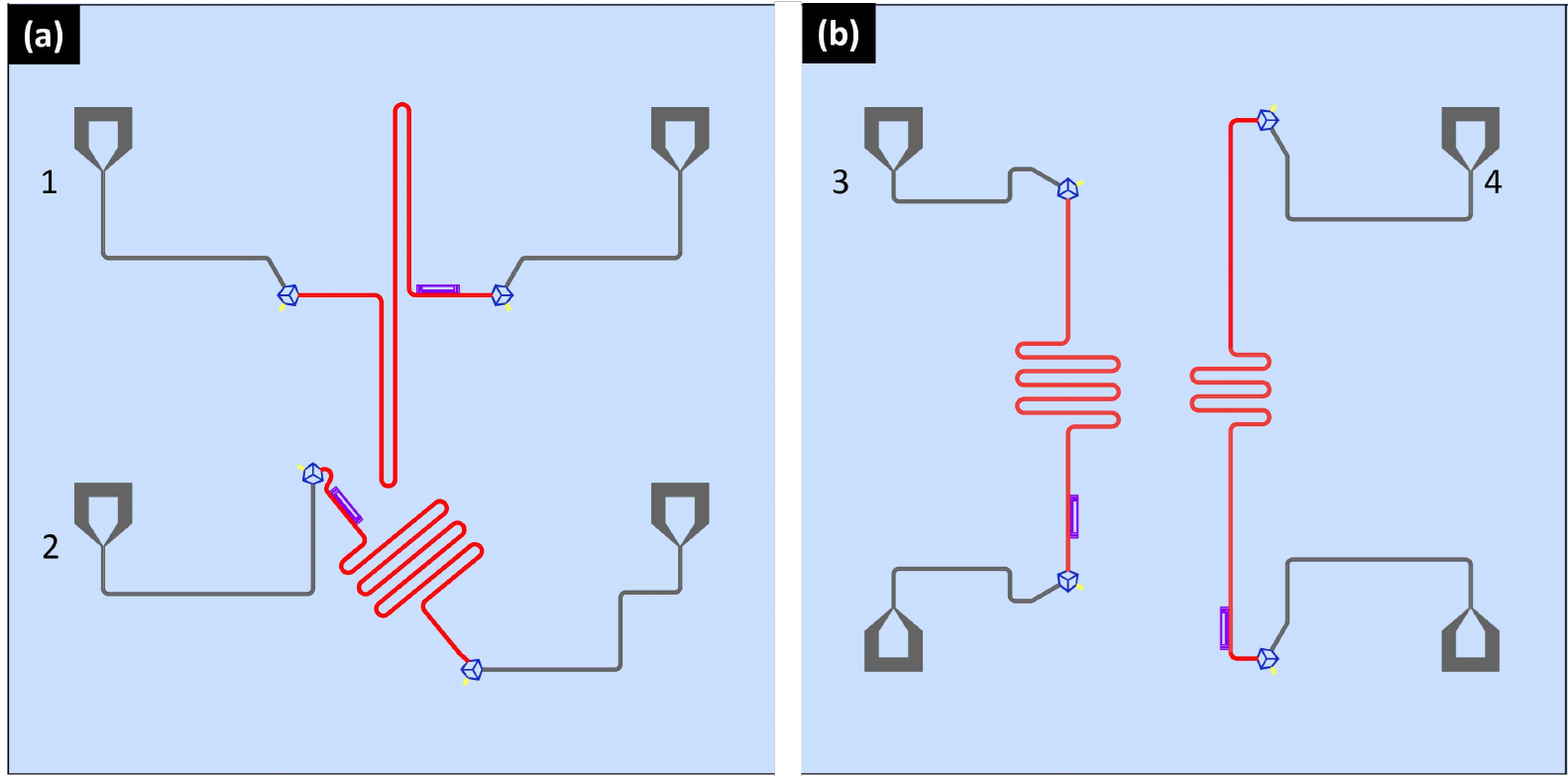}
    \vspace{-0.2cm}
	\caption{\label{fig:res_test} 
    \textit{Resonator Shape Test.} 
    (a)-(b) CAD of the two shape-test devices that were fabricated and measured to determine the frequencies of the four distinct CPW resonator shapes used. The CPW resonators, shown in red, are each capacitively coupled to an input and an output line, shown in grey. The couplers are shown in dark blue. Two capacitor pads, highlighted in purple, are coupled to the side of the resonator to account for the shift in the frequency of the resonator from adding a transmon. High-frequency $\lambda/4$ resonators, used to terminate unused coupler parts, are highlighted in yellow. Resonator $1$ is found to have a significantly lower frequency than the other three resonators. This effect is compensated for in the full resonator array by removing $45.48 \ \mu$m from all resonators with the same shape as Resonator $1$.
    }  
\end{figure*}

\subsection{On-Site Energies}\label{app_subsec:ShapeTest}

The resonator array featured in this work is composed of nine identical unit cells, each containing six CPW resonators.
Each resonator is designed to have the first harmonic at $5$ GHz and the second at $10$ GHz, and unintended frequency differences between resonators would introduce disorder to the lattice which in turn can result in undesired localization.
As such, the designs of the resonators utilized in our final device were intentionally chosen to minimize frequency disorder.
More specifically, we use CPWs with wide center pins and gentle bend radii that are larger than the field standard to reduce sensitivity to disorder introduced during device fabrication \cite{Houck_Nature_QS}. 
Our CPWs have a center-pin width of $20$~$\mu$m, a gap width of $8.372$~$\mu$m, and bend radius of $90$~$\mu$m.

Physically realizing the quasi-1D target lattice on a square chip requires the usage of multiple resonator shapes throughout the device.
The resonant frequency of a CPW is primarily set by the length of the resonator, and to leading order, the choice in resonator shape does not introduce frequency disorder.
However, higher-order corrections to the resonant frequency can arise, e.g. due to the number of bends in the resonator and variations in the bend radii.
To minimize the impact of frequency disorder between resonator shapes and the amount of calibration needed, we only use four distinct resonator shapes and we correct systematic frequency differences between them by adding or subtracting from the length of the CPW.
In order to determine the corrections needed, we tested each resonator geometry individually.
The CAD of the two test chips is shown in Fig.~\ref{fig:res_test}.
In order to separate geometry-dependent frequency shifts from the effects of run-to-run variations in lithography exposure, both devices were exposed and etched on a single chip before being diced and separated.
The half-wave frequency of the resonator with the fewest bends (Resonator 1 in Fig.~\ref{fig:res_test}(a)) differed from the other three designs by nearly $20$ MHz: Resonator 1 had a frequency of $4.8778$~GHz while Resonators 2-4 had frequencies of $4.8957 \pm 0.001$~GHz. 
For the final design, a length of $45.48 \ \mu$m was removed from Resonator 1 to match its frequency with the rest of the resonators. 

In Section \ref{subsec:modespec}, the characteristic frequency of the half-wave mode for the full resonator lattice is found to be $2\pi \times 4.889$~GHz by comparing mode-mode spectroscopy of the half-wave modes with the density of states model for the target lattice, $\sim 2\pi \times 10$~MHz off from the measured individual resonator frequency, and is likely due to slight exposure differences between the test run and the final device.

Another potential source of disorder arises from the capacitance between neighboring resonators, since the coupling capacitors on the device, highlighted blue in Fig.~\ref{fig:res_test}, induce on-site energy shifts in the resonators.
To keep this contribution uniform, all resonators must have the same number of coupling capacitors \cite{Koch:TRS_breaking, Kollar:2019hyperbolic}, two at each end for the case of the device shown here. 
Therefore, the same 3-way coupling-capacitor geometry is used at the ends of all resonators, both for sites in the bulk which naturally have four neighbors, as well as for sites at the edge of the device, where some of the neighboring resonators are absent. 
Unused ports on the 3-way capacitors are either connected to input/output feedlines, or terminated with a high-frequency $\lambda/4$ resonator, which does not participate in the lattice due to the large detuning.

The capacitor paddles of a transmon, highlighted in purple in Fig.~\ref{fig:res_test}, also modify the frequency of a CPW by introducing a significant change in the local inductance and capacitance per unit length of the resonator in which the transmon is housed. 
If not accounted for, this would cause the resonators containing the qubits to be frequency shifted compared to all other sites of the lattice. 
To eliminate this parasitic effect, identical capacitor paddles are included at every lattice site (though Josephson junctions are only written at the sites designated to have an active qubit).
The beginning of a flux bias line was also placed at every site, but these are solely included for the ease of flux bias line routing.
Since CPW resonators have standing-wave modes, the effect of the transmon capacitor depends on the position along the length of the resonator. While this could, in principle, be removed in the frequency calibration procedure described above, we constrain all four resonator shapes to have the qubit capacitor paddles located at the same distance from the end. This also helps maintain uniformity of the single-resonator $g$.

\subsection{Hopping Parameters}\label{app_subsec:HoppingParams}

To determine the predicted zeroth-order hopping rate in our device, we use the resonant frequencies determined from the shape-test devices (or finite element simulations in HFSS), $Z_0 = 50\ \Omega$, and a finite element simulation of the coupling capacitance using Ansys Maxwell.
%That leaves the end-to-end capacitance of the resonators, which is obtained with a finite element simulation in Ansys Maxwell. 
Our three-way coupler with a gap of $2~\mu m$ is simulated to have a $19.5$~fF capacitance between each pair of resonators, from which
$\vert t_{1}^{(0)} \vert/ 2\pi$ is predicted to be $\sim 47$~MHz and $\vert t_{2}^{(0)} \vert/ 2\pi$ is predicted to be $\sim 94$~MHz. 
For the main device, the empirically determined hopping rates for the half-wave and full-wave modes are determined through comparison of the experimentally-measured photonic bands with the frequency-dependent hopping model of the target lattice (see Section~\ref{subsec:modespec}). These values of $\vert t_{1}^{(0)} \vert/ 2\pi \approx 40$~MHz and $\vert t_{2}^{(0)} \vert/ 2\pi \approx 82$~MHz are reasonably close to the simulated values predicted by circuit theory and finite-element simulation. The discrepancy may be due to slight over-exposure of the device during photolithography, or due to inaccuracy in the Ansys Maxwell capacitance simulation at the level of $\sim 1$~fF.

\subsection{Transmon Design}\label{app_subsec:Transmon}

%%%%% second copy of table, double column
\begin{table*}
\caption{\label{app:paramtable}%
\textit{Device Design Parameters.} Comparison of numerically simulated device parameters and measured values for the inter-CPW hopping, transmons, and qubit-resonator coupling.  }
\begin{ruledtabular}
\def\arraystretch{1.2}%
\setlength{\tabcolsep}{4pt}
\begin{tabular}{cccc}
Resonator & $\vert t_{1} \vert/2\pi$ & $\vert t_{2} \vert/2\pi$\\
\hline
Finite Element & $47$ \mbox{MHz} & $94$ \mbox{MHz}\\
DOS Comparison & $40$ MHz & $82$ \mbox{MHz} \\
\hline
Transmon & $E_C/ 2\pi$ & $E_{J,sum}/ 2\pi$\\
\hline
Finite Element & $165$ \mbox{MHz} & - \\
\sixteen ~(Measured) & $125$ \mbox{MHz} & $100$ \mbox{GHz} \\
\five ~(Measured) & $113$ \mbox{MHz} & $104$ \mbox{GHz} \\
\eight ~(Measured) & $122$ \mbox{MHz} & $103$ \mbox{GHz} \\
\hline
Qubit-Resonator Coupling & $g_1/ 2\pi$ & $g_2/ 2\pi$ \\
\hline
Finite Element & $76$ \mbox{MHz} & $152$ \mbox{MHz} \\
Measurement & $82.5$ \mbox{MHz} & $165$ \mbox{MHz} \\
\end{tabular}
\end{ruledtabular}
\end{table*}

Before the design of the transmon qubits embedded in the CPW lattice is discussed in detail, we first introduce some of the traditional parameters of a transmon, as defined in Refs. \cite{Koch:transmon,Blais:revmodphys,quantumengineersguide}. 
The frequency of the transition between the ground and first-excited state is approximately given by 
\begin{equation}\label{eq:transmon_freq}
    \omega_{01} = \sqrt{8E_J E_C} - E_C,
\end{equation}
where $E_J$ is the Josephson energy and $E_C$ is the capacitive charging energy. The parameter $E_C$ is generally fixed, dependent on the total capacitance between the two qubit paddles, $C_{\Sigma}$:
\begin{equation}\label{eq:E_C}
E_C = \frac{e^2}{2 C_{\Sigma}}.
\end{equation}
The total capacitance takes into account not only the direct capacitance between the two paddles but also the indirect coupling through the center pin of the CPW resonator and the ground planes of the device \cite{quantumengineersguide}. For flux-tunable transmons, the effective total $E_J$ is dependent on the individual Josephson energies of the two junctions in the SQUID loop and the external applied flux. The full form for the effective $E_J$ of a tunable transmon is given by
% \begin{equation}\label{eq:E_J}
% E_J = E_{J, sum} \cos \left( \frac{\pi \Phi_{ext}}{\Phi_0} \right) \sqrt{1+\left( \frac{E_{J,diff}}{E_{J,sum}} \right)^2 \tan^2 \left( \frac{\pi \Phi_{ext}}{\Phi_0} \right)},
% \end{equation}
% \begin{eqnarray}\label{eq:E_J}
% E_J = E_{J, sum} & \cos \left( \frac{\pi \Phi_{ext}}{\Phi_0} \right) \\ \nonumber
%  & \times \sqrt{1+\left( \frac{E_{J,diff}}{E_{J,sum}} \right)^2 \tan^2 \left( \frac{\pi \Phi_{ext}}{\Phi_0} \right)},
% \end{eqnarray}
\begin{multline}\label{eq:E_J}
E_J = E_{J, sum} \cos \left( \frac{\pi \Phi_{ext}}{\Phi_0} \right) \\ 
 \times \sqrt{1+\left( \frac{E_{J,diff}}{E_{J,sum}} \right)^2 \tan^2 \left( \frac{\pi \Phi_{ext}}{\Phi_0} \right)},
\end{multline}
where $\Phi_0$ is the superconducting magnetic flux quantum, $E_{J,sum} = E_{J1} + E_{J2}$ is the sum of the two individual Josephson energies, and $E_{J,diff} = \vert E_{J1} - E_{J2} \vert$ is the difference between the two individual Josephson energies. 
The effective Josephson energy (and consequently the qubit frequency) can be tuned by varying the external magnetic flux $\Phi_{ext}$ through the SQUID loop.

In order to suppress the effect of charge noise in the substrate, a transmon must operate in the regime where $E_J/E_C \gg 1$, typically $20$ or larger \cite{Koch:transmon, Blais:revmodphys}.
This requires a large $C_{\Sigma}$, which can either be achieved using large-area capacitor paddles or a digitated capacitor.
Our design uses a digitated capacitor in order to make the qubit compact and easier to incorporate in different resonator geometries.
Conventional, non-digitated designs would need to be significantly wider, as much as $5-10$ times, to achieve an equivalent capacitance, which would severely restrict the ability to tightly pack resonators in the array \cite{Gambetta:fab}.
For our design, the total capacitance between the paddles is determined to be $C_{\Sigma} = 0.1174$~pF using Ansys Maxwell.
Following from Eq. \ref{eq:E_C}, $E_C/ 2\pi$ is predicted to be $165$ MHz. 
In a transmon, the two-photon transition from the ground state to the second-excited state is lower in frequency than the main transition by $\sim E_C/2$ \cite{Blais:revmodphys, quantumengineersguide}. 
As a result, measuring this multiphoton transition using two-tone spectroscopy allows $E_C$ to be determined in situ for each qubit.
For the three qubits, we find $E_C/ 2\pi = 125$~MHz, $113$~MHz, and $122$~MHz, respectively.
%$2\pi \times 122$ MHz in average.
%\wcl{16: 123.5, 8: 120.5}

The junctions that form the SQUID loop of the transmon are deposited via double-angle evaporation (see Appendix \ref{app:Fab} for a detailed description of the fabrication process).
As the junctions for the three transmons are fabricated simultaneously, the junctions must all share the same orientation.
Thus, while the capacitor paddles are always parallel to the center pin of the coupled resonator, the junctions are instead parallel to the edge of the chip, leading to a minor difference in the designs for the on-axis qubit (\five) and the off-axis qubits (\sixteen ~and \eight).
The transmons are designed to have tuning ranges that span from below the half-wave modes up to the full-wave modes.
To experimentally determine $E_{J,sum}$ of the fabricated junctions, we match Eq.~\ref{eq:transmon_freq} and Eq.~\ref{eq:E_J} to two-tone spectroscopy scans that map out the flux dependence of the transmon frequency and the photonic bound states (see Appendix \ref{app:gfit} for details). For the three qubits, we find $E_{J,sum}/ 2\pi = 100$~GHz, $104$~GHz, and $103$~GHz, respectively.

A summary of the simulated and empirically-determined device parameters presented in this section is shown in Table \ref{app:paramtable}.

\section{Determination of $g$}\label{app:gfit}

\begin{figure*}[ht!]
\centering
    \includegraphics[width=1\textwidth]{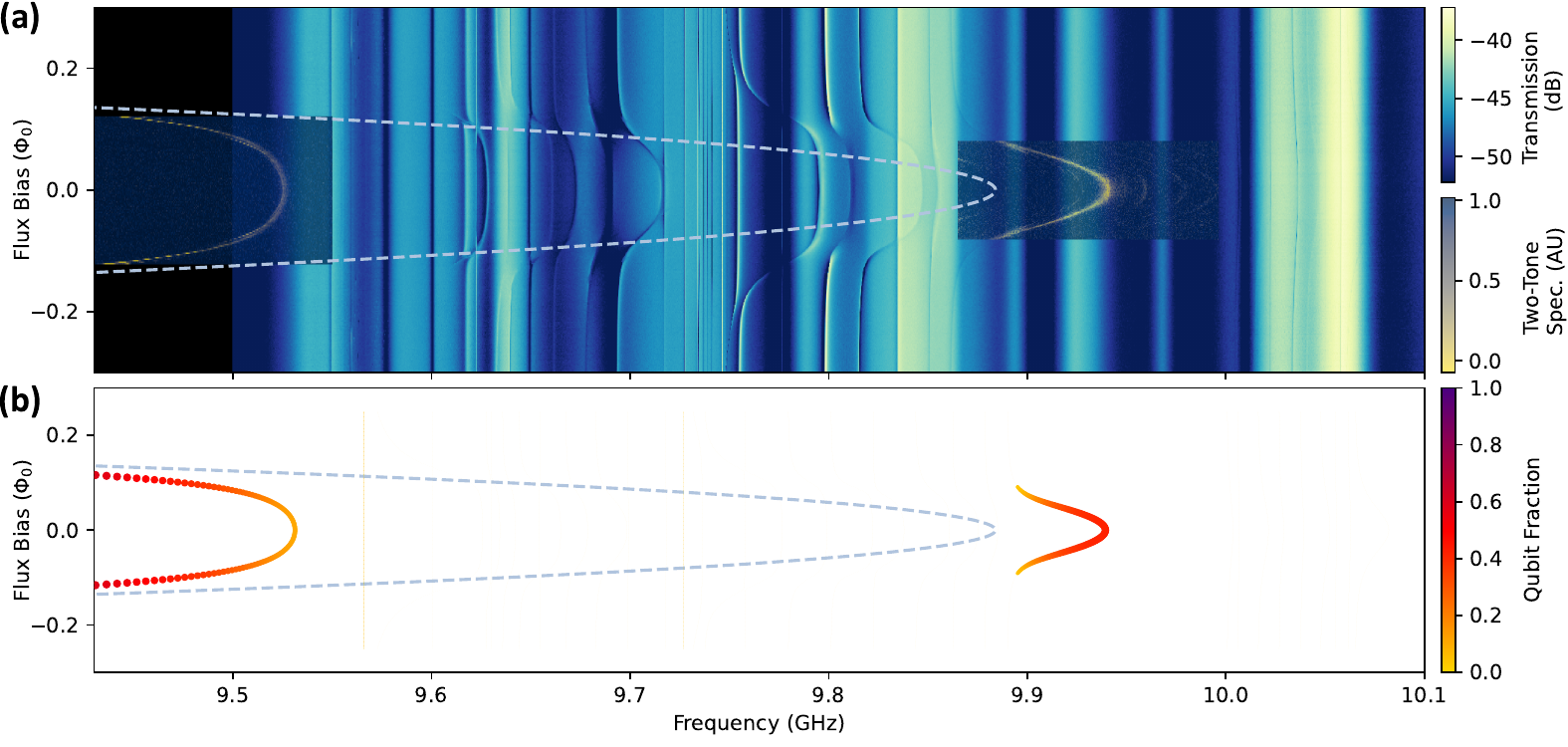}
    \vspace{-0.2cm}
	\caption{\label{fig:gfig} 
    \textit{Comparison of $g$ and the photonic bound states.} 
    (a) Transmission and two-tone qubit-spectroscopy data near the full-wave modes of the lattice. (Reproduced from Fig.~\ref{fig:FWdata}) for \eight. The dashed line indicates the location of the bare-qubit transition used for the theoretical calculation in (b). (b) Theoretical calculation of the location of photonic bound states (red/orange) versus flux  for a transmon (approximated as flux-tunable two-level system) coupled to the lattice with a coupling strength $g_2/ 2\pi = 165$~MHz. 
    The size and color of the marker indicates the fraction of the wave function which is qubit-like. Photon-like eigenstates not shown.
    The grey dashed line indicates the flux dependence of the bare qubit transition.
    The locations of the photonic bound states in (a) and (b) are in good agreement, indicating that qubits in our device have single-cavity coupling strengths $g_2/ 2\pi \approx 165$~MHz and $g_1/ 2\pi = (g_2/2)/ 2\pi \approx 82.5$~MHz.
    }  
\end{figure*}

%\ak{to add: what is the traditional method more explicitly. Option 1: measure vacuum rabi splitting from sweeping through. Option 2: dispersive shift/photon number splitting. Reword for clarity/emphasis.}

In a conventional single-mode circuit QED system, the qubit-cavity coupling strength $g_0$ could be determined by measuring the vacuum Rabi splitting when the qubit frequency is swept through resonance with the mode \cite{Blais:revmodphys}.
Alternatively, the coupling could be determined from the dispersive shifts (or Stark shifts) on the qubit due to a detuned cavity.
However, in our device, both of these methods can be problematic.
For the resonant case, avoided crossings between the qubits and the normal modes of the lattice are clearly visible in Figs. \ref{fig:HWwide}, \ref{fig:FWwide}, \ref{fig:HWzooms}, and \ref{fig:FWzooms}. 
However, the lattice modes are close enough together in frequency that these avoided crossings are not independent.
The Rabi split peaks from the qubit interacting with mode $k$ can undergo avoided crossings with those from mode $k+1$ and hide the true size of the splitting. 
These avoided crossings, therefore, cannot be directly used to reliably extract the normal mode couplings $g_{k,j}$. 
Additionally, the normal-mode couplings $g_{k,j}$ cannot be used to determine the single-cavity coupling $g_0$ without detailed knowledge of the normal-mode wave functions in the presence of disorder.

For the far-detuned case, the photon-number dependent dispersive shift, or AC Stark shift, on qubit $j$ scales as $n*g_{k,j}^2$, where $n$ is the intracavity microwave photon number in mode $k$ \cite{Blais:revmodphys}.
Provided a strong enough qubit-cavity coupling, this shift produces resolvable peaks in the qubit transition frequency for each value of the intracavity photon number; the frequency difference between the transitions for $n$ photons and $n+1$ photons can subsequently be used to extract $g_0$ \cite{Schuster_2007:Number_Splitting}.
In our device, the normal-mode couplings involved are small enough that the Stark shifts on the qubits are not single-photon resolved and therefore cannot be used to calibrate $g_{k,j}$ without an independent measure of the intracavity photon number.

As a result, the most reliable way to determine $g_0$ is from the photonic bound states. Fig.~\ref{fig:gfig} shows a comparison between the experimental photonic-bound-state data (first shown in Fig.~\ref{fig:FWdata}) and a numerical simulation of the photonic bound states. In order to simulate the photonic bound state, we numerically diagonalize the single-excitation manifold of the multimode Jaynes-Cummings Hamiltonian in Eq.~\ref{eqn:multimodeJCint}. 
We approximate the transmon as a two-level system, neglecting all higher levels, and compute the lattice modes for a $9$-unit cell chain with hard-wall boundary conditions, treated within the self-consistent tight-binding approximation using the best-fit hopping parameters given in Table~\ref{app:paramtable} (extracted from Figs.~\ref{fig:HWdata}
and \ref{fig:FWdata}).

Good qualitative agreement between the experimentally-measured and the numerically-simulated photonic bound states is found for 
$g_2/ 2\pi = 165$~MHz at the full-wave modes. The corresponding coupling strength at the half-wave modes is then inferred to be $g_1/ 2\pi = (g_2/2)/ 2\pi = 82.5$~MHz. These values are close to the values of $g_1/ 2\pi = 76$~MHz and $g_2/ 2\pi = 152$~MHz derived from finite element simulations. The discrepancy is likely due to slight over or under-exposure of the design during fabrication. 

A more quantitative technique to extract $g_{k,j}$ from non-linear Kerr spectroscopy is the subject of upcoming work in Ref.~\cite{KerrSpec}.
%However, in our device the dispersive shift from the lattice modes is not strong enough to resolve the discrete photon number states, meaning this is also not a viable method to determine mode coupling.
%-unlike single-mode systems where extracting g from photon number splitting is standard. The normal-mode gs here are smaller. Not single photon resolved. 
%A more quantitative technique to extract $g_{k,j}$ from non-linear spectroscopy is the subject of upcoming work in Ref.~\cite{XYZ}. 

\section{Transmon Ionization and Mode-Mode Spectroscopy}
\label{app:ionization}

In Section~\ref{subsec:modespec}, we introduced the measurement technique mode-mode spectroscopy in which lattice modes can be detected with high SNR using their qubit-induced nonlinearity.
In this method the response of a strongly driven mode is detected via the frequency shift induced on a weakly-driven monitor mode.
At low drive powers, the signal originates from Kerr nonlinearity \cite{Peugeot_2024,Bosman_2017}.
However, the mode-mode spectroscopy presented in this work is collected at comparatively high drive powers, where the frequency shift of the monitor mode primarily arises from transmon ionization.

Transmon ionization is a process in which a transmon is excited above its cosine potential well. In conventional transmon readout, this process leads to measurement errors \cite{Blais:ionization,Martinis:ionization}.
When the drive amplitude applied to the resonator is high enough, the transmon can be excited directly to higher energy states through a multiphoton process, even with a far off-resonant drive.
Here, the same principle is exploited to perform high-power mode-mode spectroscopy on our device.
We apply a scanning drive tone which does not affect the transmon when off resonance, but when it comes into resonance with one of the photonic modes of the lattice, the resulting intracavity photon number is large enough to ionize one or more of the transmons.
The resulting frequency shifts of the monitor mode, and associated transmission changes, are \emph{much} larger than what is observed at low drive power, due to the lowest-order Kerr non-linearity alone.

\section{Flux calibration}\label{app:FluxCal}

In order to carry out all of the measurements above with the three qubits at well-known frequencies, a feed-forward scheme was implemented to remove cross-talk between the on-chip flux bias lines, following the procedure outlined in Ref.~\cite{Ma:2019ba}. 
By first measuring and then inverting the cross-talk between the lines, a software correction can be applied in order to tune each qubit independently. 
This allows individual qubits to be tuned to well-determined detunings from the different photonic bands.

Assuming that the flux through each SQUID loop ($\phi_i$) responds linearly to the voltage applied to the flux bias lines ($V_i$), the relationship between the voltage and flux is described by the following equation: 

\begin{equation}\label{eq:v2f}
    \begin{pmatrix} \phi_1 \\ \phi_2 \\
    \phi_3 \end{pmatrix} = M \begin{pmatrix} V_1 \\ V_2 \\ V_3 \end{pmatrix} + \begin{pmatrix} \phi_1^* \\ \phi_2^* \\ \phi_3^* \end{pmatrix}
\end{equation}
Here, the cross-talk matrix $M$ has elements $M_{ij} = \partial \phi_i / \partial V_j$, and $\phi_i^*$ is the flux offset at zero voltage bias for qubit $i$. Once $M$ and the flux offsets are identified, Equation \ref{eq:v2f} can be solved for the voltages necessary to target any desired flux vector. 

The first step is to find the volts per flux quantum for each qubit, which gives the diagonal elements of the matrix $M_{ii}$. To accomplish this, we select a cavity mode to monitor in transmission and perform sweeps of $V_i$ for each flux bias line, finding where qubit $i$ crosses that mode. Since these crossings occur periodically in flux, we can extract both the volts per flux quantum ($M_{ii}$) and the flux offset ($\phi_i^*$).

The next and final step is to determine the off-diagonal elements. Typically, these off-diagonal elements are too small for a direct measurement of the flux quantum to be feasible.
Instead, we extract them from two-tone spectroscopy scans.
One of the qubits is parked at a frequency $\omega_i^*$ and the applied voltage to one of the lines ($V_j$) is swept while monitoring the qubit frequency ($\omega_i$).
The voltage range swept is small enough for the frequency response to be roughly linear and 
%$\partial \omega_i / \partial V_j|^{\omega_i^*}$
$\tfrac{\partial \omega_i }{ \partial V_j} \big|_{\omega_i^*}$
is extracted from the slope of the scan. 
This measurement is repeated for all possible pairs of qubits and flux bias lines, nine times in total.
Using $\tfrac{\partial \phi_i }{ \partial V_i}$ from the transmission scans and $\tfrac{\partial \omega_i }{ \partial V_i} \big|_{\omega_i^*}$ from the two-tone spectroscopy scans, we compute $\tfrac{\partial \omega_i }{ \partial \phi_i} \big|_{\omega_i^*}$, the slope with respect to the applied flux at this particular frequency point. 
We then divide $\tfrac{\partial \omega_i }{ \partial V_j} \big|_{\omega_i^*}$ by $\tfrac{\partial \omega_i }{ \partial \phi_i} \big|_{\omega_i^*}$ to get $\tfrac{\partial \phi_i }{ \partial V_j}$, which yields the remaining off-diagonal cross-talk matrix elements $M_{ij}$.

% \begin{figure*}[ht!]
% \centering
%     \includegraphics[width=1\textwidth]{Figures/FM8_g_MatchingFig.pdf}
%     \vspace{-0.2cm}
% 	\caption{\label{fig:gfig} 
%     \textit{Comparison of $g$ and the photonic bound states.} 
%     (a) Transmission and two-tone qubit-spectroscopy data near the full-wave modes of the lattice. (Reproduced from Fig.~\ref{fig:FWdata}) for \eight. The dashed line indicates the location of the bare-qubit transition used for the theoretical calculation in (b). (b) Theoretical calculation of the location of photonic bound states (red/orange) versus flux  for a transmon (approximated as flux-tunable two-level system) coupled to the lattice with a coupling strength $g_2/ 2\pi = 165$~MHz. 
%     The size and color of the marker indicates the fraction of the wave function which is qubit-like. Photon-like eigenstates not shown.
%     The grey dashed line indicates the flux dependence of the bare qubit transition.
%     The locations of the photonic bound states in (a) and (b) are in good agreement, indicating that qubits in our device have single-cavity coupling strengths $g_2/ 2\pi \approx 165$~MHz and $g_1/ 2\pi = (g_2/2)/ 2\pi \approx 82.5$~MHz.
%     }  
% \end{figure*}

\begin{figure*}[t]
\centering
    \includegraphics[width=5in]{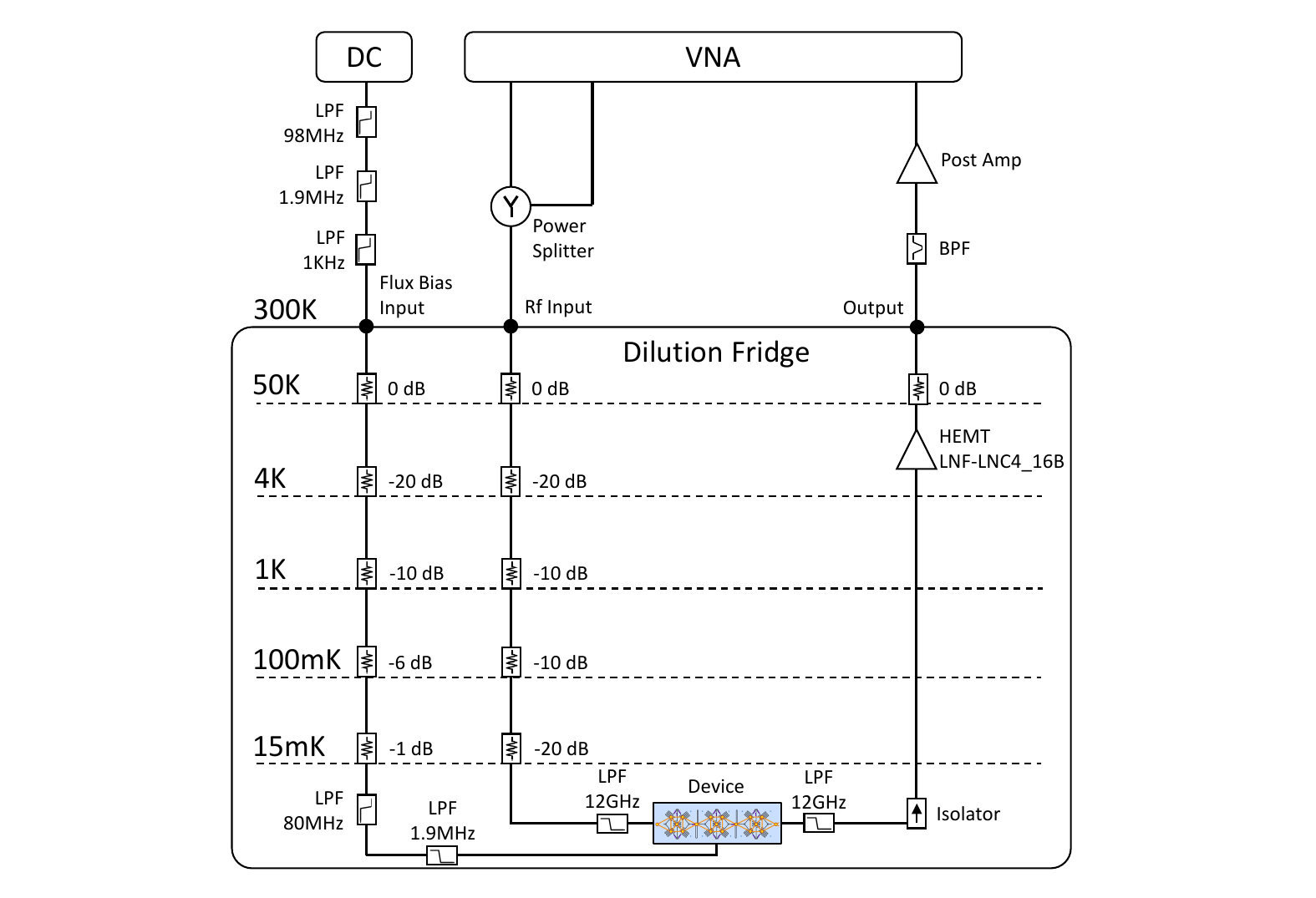}
    \vspace{-0.2cm}
    \caption{\label{fig:LineSchematic} 
    \textit{Experimental setup.} 
    Schematic for the RF lines for readout and qubit control and the DC lines used to flux-bias the qubits. RF signals from a vector network analyzer (VNA) undergo $-60$~dB nominal attenuation inside the dilution fridge and pass through a $12$~GHz lowpass filter (LPF) prior to reaching the device. The output RF signals from the device pass through a second $12$~GHz LPF, then an isolator, and are then amplified by a high electron mobility transistor  (HEMT) amplifier. Outside the fridge, the signals pass through a bandpass filter (BPF, $3.4$~GHz \textendash ~$5.5$~GHz for measurements of the half-wave modes and $9$~GHz \textendash ~$12$~GHz for measurements of the full-wave modes) and are amplified further before entering a readout port on the VNA. The DC voltage applied to each flux-bias line passes through three LPFs outside of the fridge, undergoes $-37$~dB nominal attenuation inside the fridge and passes through two more LPFs before reaching the device. 
    %A VNA and DC power supplies are located outside of the dilution refrigerator. The VNA is used to generate input signals, which are then combined with a power splitter. The DC power supplies are used for voltage biasing to generate magnetic flux in the device. Both inputs undergo a series of attenuations before reaching the device. The output signal is protected by an isolator and amplified by a HEMT and a post-amplifier chain.
    }
\end{figure*}
%%%%%%%%%%%%%%%%%%%%%%%%%%%%%%%%%%%%%%%%%

%%%%%%%%%%%%%%%%%%%%%%%%%%%%%%%%%%
\section{Device Fabrication}\label{app:Fab}

The lattice device presented in this work was fabricated on a $25.4 \times 25.4$~mm$^2$ sapphire wafer coated with a $200$~nm film of tantalum.
The fabrication process was carried out in three main steps: photolithography, electron-beam (e-beam) lithography, and double-angle evaporation.
The larger features, including the resonators, couplers, and qubit capacitor paddles, were fabricated using photolithography, while the SQUID loops for the transmons were defined with electron-beam lithography. In the double-angle evaporation step, aluminum was evaporated onto the chip to form the SQUID loops.

The photolithography was carried out using the Heidelberg MLA150 direct write system. 
Prior to exposure with the Heidelberg, the chip was cleaned with Remover PG, acetone, and isopropanol in sequence before being coated with photoresist (AZ 1518).
In the process of spinning on the photoresist, more resist gathers near the edges of the chip than near the center (also known as the edge bead effect). 
To compensate for these edge beads, the edges of the chip were exposed at a higher power than the rest of the chip.  
The three-way coupler connecting the input port of the device to the first resonator in the chain is closer to the edge of the chip than all other couplers and was also exposed with a higher dosage to compensate for edge beads.
After the exposure, the chip was developed in MF CD-26 and the pattern was etched using Tantalum Etchant 111 from the Transene Company. The chip was then cleaned again with Remover PG, acetone, and isopropanol

The SQUID loops were written using an Elionix ELS-G100 electron-beam lithography system. To prepare for e-beam lithography, the chip was cleaned with toluene, acetone, methanol, and isopropanol in sequence. Next, the chip was cleaned with piranha solution to remove organic residues \cite{Piranha_etch_deleon}. Two layers of e-beam resist, MMA (MMA EL13) and PMMA (MMA 950 A3), were then spun onto the chip. Having two resist layers allows for the addition of undercuts in the lithography process. The sample was covered with a 30 nm anticharging layer of aluminum and then placed in the Elionix, where Manhattan-style junctions were defined through e-beam lithography. After the write, the anti-charging layer was removed with AZ 300 MIF and the sample was developed in MIBK:IPA (1:3). At this stage, the remaining e-beam resist on the chip forms channels for evaporated aluminum to fill in order to create the desired SQUID loops.

Lastly, we performed double-angle evaporation with a Plassys MEB 550S.
The Josephson junctions for all three qubits were designed to require the same two deposition angles to simplify the fabrication process.
We deposited a 30 nm layer of aluminum along one angle, oxidized the sample, then deposited a 50 nm layer of aluminum along the second angle.
Lift-off of the excess aluminum was performed using Remover PG and isopropanol. Finally, the chip was connected to a copper-enclosured printed circuit board through wire bonding. Wire bonds were also used to connect the ground planes across all resonators.

\section{Measurement Scheme}\label{app:meas_scheme}

Our quasi-1D lattice device is measured with the continuous wave (CW) measurement scheme depicted in Fig.~\ref{fig:LineSchematic}. A vector network analyzer (VNA, Rohde-Schwarz ZNB20) is used to generate CW signals from one or two separated ports. The two output signals are combined with a microwave power splitter (ZFRSC-183-S+) and undergo $-60$ dB nominal attenuation inside the dilution fridge before reaching the device. The input and output of the device are protected by two identical low pass filters (LPF, K$\&$L 6L250-00089) with a cut-off frequency of $12$~GHz. An isolator (QCI-G0401202AS) on the base plate protects the device from reflected signals. The signal is then amplified by a high electron mobility transistor (HEMT) amplifier (LNF-LNC4\_16B), which is specially chosen to amplify signals in the range of $4$\textendash $16$~GHz encompassing both the half-wave and the full-wave modes. Finally, a post amplifier chain is located outside of the dilution fridge to amplify the signal further before it is sent into a readout port on the VNA. A band pass filter (BPF) is set before the post-amplifiers to reduce noise outside of the desired bandwidth. The frequency range chosen changes between measurements of the half-wave (3.4 GHz \textendash ~5.5 GHz) and full-wave (9 GHz \textendash ~12 GHz) modes. 

DC voltage sources (Stanford Research Systems DC205) are used to flux bias the qubits. There are three identical flux bias lines (one for each qubit) but for simplicity only one is shown in Fig.~\ref{fig:LineSchematic}. The DC voltage passes through a series of low pass filters and attenuators before reaching the device, where the current generates a magnetic flux through the SQUID loop of the transmon.

\bibliographystyle{apsrev4-2}
\bibliography{refs.bib}

\end{document}